\documentclass[preprint,12pt,3p]{elsarticle}
\usepackage{lineno,hyperref}
\modulolinenumbers[5]

\journal{Acta Materialia}

\usepackage{graphicx}
\usepackage{amsmath}
\usepackage{booktabs}
\usepackage{subfig}
\usepackage{caption}
\usepackage{float}
\usepackage{epstopdf}
\usepackage[version=3]{mhchem} 	
\graphicspath{{./pics/}}
\usepackage{color,soul}

\begin{document}

\begin{frontmatter}

\title{Modeling the $\alpha/\omega$ Thermal Stability in Shocked Zr: A Coupling between Dislocation Removal and Phase Transformation}

\author[l1]{T.S.E. Low}
\author[l1,l2]{S.R. Niezgoda\corref{cor1}} 
\cortext[cor1]{Corresponding Author}
\ead{niezgoda.6@osu.edu}

\address[l1]{Department of Materials Science and Engineering, The Ohio State University, Columbus, OH 43210, USA}
\address[l2]{Department of Mechanical and Aerospace Engineering, The Ohio State University, Columbus, OH 43210, USA}

\begin{abstract}
Under high pressure, Zr undergoes a transformation from its ambient equilibrium hexagonal close packed $\alpha$ phase to a simple hexagonal $\omega$ phase. Subsequent unloading to ambient conditions does not see a full reversal to the $\alpha$ phase, but rather a retainment of significant $\omega$. Previously, the thermal stability of the $\omega$ phase was investigated via in-situ synchrotron X-ray diffraction analysis of the isothermal annealing of Zr samples shocked to 8 and 10.5 GPa at temperatures 443, 463, 483, and 503 K. The phase volume fractions were tracked quantitatively and the dislocation densities were tracked semi-quantitatively. Trends included a rapid initial (transient) transformation rate from $\omega\to\alpha$ followed by a plateau to a new metastable state with lesser retained $\omega$ (asymptotic). A significant reduction in dislocation densities in the $\omega$ phase was observed prior to initiation of an earnest reverse transformation, leading to the hypothesis that the $\omega\to\alpha$ transformation from is being hindered by defects in the $\omega$ phase. As a continuation of this work, we present a temperature dependent model that couples the removal of dislocations in the $\omega$ phase and the reverse transformation via a barrier energy that is associated with the free energy of remaining dislocations. The reduction of dislocations in the $\omega$ phase occurs as a sum of glide and climb controlled processes, both of which dictate the transient and asymptotic behavior of the annealing process respectively. 
\end{abstract}


\end{frontmatter}

\section{Introduction}
At ambient conditions, the group IV transition metals (Zr, Ti, Hf) are thermodynamically stable as a hexagonal close packed (HCP) crystal ($P6_3/mmc$, $c/a=1.593$), denoted as the $\alpha$ phase. At high pressure, the $\alpha$ phase in these metals undergoes a transformation to the $\omega$ phase, which has a simple hexagonal crystal structure ($P6/mmm$, $c/a=0.623$). After shock loading \cite{Song1994} or severe plastic deformation by high-pressure torsion (HPT) \cite{perez2008bulk, edalati2009}, the $\omega$ phase displays a strong hysteresis; with a significant fraction remaining after pressure is released. In the case of $\omega$ material yielded from shock impact experiments, samples with retained $\omega$ fractions as high as  80\% of $\omega$ phase have been observed \cite{cerreta2005influence, cerreta2012influence, Cerreta20137712}. The resulting two-phase microstructure is (meta)stable for years at standard temperature and pressure. 

Over the past 60 years, there have been numerous studies on the $\alpha\to\omega$ forward transformation, with the majority focusing on the determination of the equilibrium temperature/pressure phase diagram, crystallography, mechanical properties, and electronic structure of the $\omega$ phase \cite{Rabinkin, greeff2004modeling, cerreta2003shock, trinkle2005systematic}. The significant disagreement amongst these studies attest to the influence of the transformation hysteresis on the determination of basic physical properties of group IV transition metals. As an example, the equilibrium transformation pressure for Zr has been reported to range from 2--7  GPa for the static pressure loading case \cite{jamieson1963, sikka1982, jamieson1973, young1991phase}. However, even though the metastability of the $\omega$ phase is well documented, there is only a limited body of work on the stability of the $\alpha/\omega$ dual phase microstructure. 

The $\alpha\to\omega$ phase transformation in Zr has been shown to be achievable by several techniques, namely 1) hydrostatic pressure, 2) shock loading, and 3) high pressure torsion methods. It is not clear whether there is a single mechanism for the $\alpha \rightarrow \omega$ transformation or if multiple transformation pathways exist.   Depending on the experimental conditions and loading mechanisms, multiple ($\alpha/\omega$) orientation relationships have been reported. The amount of retained $\omega$, transformation pressures, and the apparent degree of $\omega$ stability vary significantly between experiments \cite{bridgman1948compression, jayaraman1963solid, olinger1973, zilbershtein1975alpha, errandonea2005}. Hence there is a need for clear need for systematic studies of both the forward transformation and of the reverse $\omega \rightarrow \alpha$ reverse transformation. More specifically, models that link the evolution of the microstructure to thermodynamic state variables, coupled with in-situ experimental observation, are required to characterize the kinetics of the transformation under various conditions in order to provide evidence for or against proposed transformation or reverse-transformation mechanisms \cite{zong2014, addessio2016, zong2016}. 

In this paper, emphasis is placed upon the \textit{thermal stability of the $\omega$ phase generated from shock experiments}. The $\omega$ phase produced from shock impact is generally characterized by the  $(0\;0\;0\;1)_\alpha\parallel(1\;0\;\overline{1}\;1)_\omega$ and $[1\;0\;\overline{1}\;0]_\alpha\parallel[1\;1\;\overline{2}\;\overline{3}]_\omega$ orientation relationship identified by Song \& Gray \cite{song1995microscopic}.  Brown et al. first investigated the thermal stability of the $\omega$ phase by heating shocked samples under a constant temperature ramp from room temperature to 620 K at a rate of 0.05 K/s \cite{Brown2014383}. A subsequent study by Low et al. involved the isothermal annealing of shocked Zr samples over a similar temperature range \cite{low2015}. Both experiments showed a significant decrease in dislocation content prior to the onset of the reverse transformation, leading to postulation that the metastability of the $\omega$ phase was highly correlated with the defect state of the microstructure. Furthermore, it was speculated that the dislocation population was somehow arresting the reverse transformation and that given sufficient thermal energy, the defects would gain mobility and annihilate through recovery mechanisms; allowing the reverse transformation to proceed. Here, we present a model based on this hypothesis and demonstrate that model predictions are consistent with the experimentally observed transformation kinetics. Section \ref{sec:methods} reviews and summarizes the previously collected experimental data which informs the current work as well as a brief discussion of  previous modeling efforts. This is followed by section \ref{section:numericalmodel} which provides detail into the model, after which section \ref{sec:results} provides the results from the model. Further discussion on the results and their implications are provided in section \ref{sec:discussion}.

\section{Experimental Methods and Prior Studies} \label{sec:methods}

\subsection{Production of Metastable $\omega$ Through Shock Impact}
\begin{figure}
	\centering
	\includegraphics[scale=0.25]{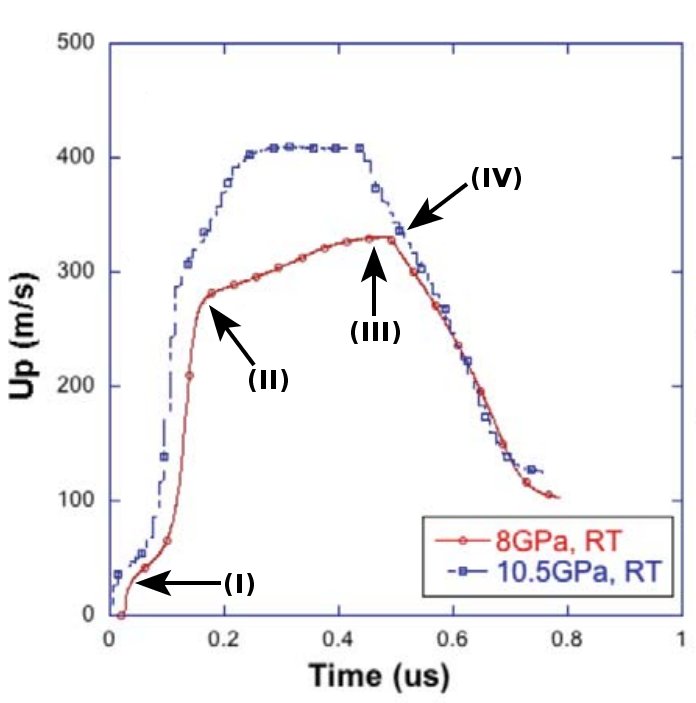}
	\caption{Wave profiles of zirconium samples shocked to 8 and 10.5 GPa at room temperature. The arrows indicate significant events during the shock impact and are interpreted as (a) Hugoniot elastic limit and onset of twin and slip deformation,  (b) onset of $\alpha\to\omega$ transformation, (c) flatlining indicating end of transformation and further deformation under peak pressure, and (d) end of shock process. Data sourced from Cerreta et al. \cite{cerreta2014}}
	\label{fig:visar}
\end{figure}

The data used to develop and validate this model came from annealing studies of shocked Zr performed by Low et al. \cite{low2015} and Brown et al. \cite{Brown2014383}.  The shocked material for those studies came from gas gun experiments performed by Ceretta and collaborators \cite{cerreta2005influence, Cerreta20137712, cerreta2003shock, cerreta2012, cerreta2006influence}. Prior to shock loading, all Zr samples were prepared from a high-purity crystal bar Zr ($<100$ ppm impurities) which was upset forged, clock rolled, and annealed, resulting in a plate with homogenous and fully recrystallized microstructure, with average grain size of 15-20 $\mu m$ \cite{escobedo2012influence}. The plate exhibited a strong basal texture ($>$8 times uniform random distribution) nearly aligned with the normal or through-thickness direction (TT) direction of the plate, with prism planes uniformly distributed about the in-plane directions of the plate (transversely isotropic). Shock loading by dynamic impact was performed utilizing a gas-driven 80-mm single stage launcher \cite{cerreta2012}. The impact direction was along the TT direction or parallel to the $c$-axis of the majority of grains in the sample. The targets were impacted by 2.5 mm thick Zr flyer plates accelerated to velocities of 640 or 835 m/s, resulting in peak compressive stresses of 8 or 10.5 GPa respectively on the Zr samples \cite{Cerreta20137712}. The total duration of the shock process (plasticity and transformation included) was approximately 0.5 $\mu$s \cite{cerreta2014}. The samples were soft-recovered and were further sectioned for X-ray and microscopy. The real-time process of the shock deformation was tracked via velocity interferometer system for any reflector (VISAR) measured wave profiles. The wave profiles and their interpretations are given in figure \ref{fig:visar}. The VISAR traces indicate that samples shocked to 10.5 GPa complete the $\alpha\to\omega$ transformation over a significantly shorter time duration than samples shocked to 8 GPa. It was also speculated that the $\omega$ phase in samples shocked to 10.5 GPa experienced additional plastic deformation, in addition to the higher pressure, resulting in a difference in defect population compared to the 8 GPa shocked samples.  The resulting volume fractions of retained $\omega$ phase are $\sim$ 60\% and $\sim$ 80\% for samples shocked to 8 and 10.5 GPa respectively. Complete details of the shock experiments can be found in \cite{ cerreta2005influence, Cerreta20137712, cerreta2003shock, cerreta2012, cerreta2006influence}. 

\subsection{In-situ Annealing Experiments}
The in-situ annealing experiments of the $\alpha/\omega$ samples consist of work by Brown et al. \cite{Brown2014383} and Low et al. \cite{low2015} and were performed via x-ray diffraction (XRD) characterization methods at the 1ID-C beam line at the Advanced Photon Source (APS), Argonne National Laboratory \cite{Haeffner2005120}. The experimental setup was typical for powder diffraction on a polycrystalline sample. A high energy monochromated incident beam (E = 86 keV) was used to illuminate the samples and the resulting diffraction images were collected on a a two-dimensional detector placed at an appropriate distance to capture at least 5 diffraction rings from each phase\cite{low2015}. The individual ($\alpha/\omega$) phase fractions were tracked quantitatively and the dislocation densities were tracked semi-quantitatively. 

The experimental work by Brown et al. involved the heating of Zr samples shocked to 8 and 10.5 GPa from 300 to 620 K at a constant temperature ramp rate of 0.05 K/s. Significant $\omega\to\alpha$ transformation was observed within the temperature range of $475 < \theta < 550 $ K  \cite{Brown2014383}.  Low et al. further investigated the transformation under isothermal conditions at temperatures of 443, 463, 483 and 503 K in samples similarly shocked to 8 and 10.5 GPa \cite{low2015} and observed a rapid initial transformation rate. At all temperatures there was a continuous deceleration of the transformation, with the transformation rate approaching zero with significant $\omega$ phase remaining. This new metastable state appeared to be asymptotically stable. The higher the temperature applied, the further the extent of the reverse transformation (see figures \ref{fig:8GPa_simulation} and \ref{fig:10GPa_simulation} in section \ref{subsection:isothermal}). Both phases contain high dislocation densities ($\approx10^{15}$ m$^{-2}$) prior to annealing and observe a reduction in dislocation densities during annealing. Furthermore, in the initial stage of annealing, a significant reduction of the $\omega$ phase dislocations was observed prior to onset of the reverse transformation, prompting the authors to postulate that the $\omega\to\alpha$ transformation was arrested by dislocations in the $\omega$ phase. The experiments also noted that at temperatures 483 and 503 K, the 10.5 GPa samples observed a more substantial reverse transformation such the $\alpha$ volume fractions surpassed those of the 8 GPa samples, despite the 10.5 GPa samples starting off with a lower $\alpha$ volume fraction. This cross-over trend was also observed when applying different temperature ramp rates from room temperature to 503 K on samples shocked to similar conditions. EBSD characterization of ex-situ annealed samples indicated that the reverse transformation proceeded primarily by growth of existing $\alpha$ laths at 443 K and below and nucleation and growth of new $\alpha$ laths at higher temperatures (463--503 K). 

The annealing experiments raised an important question. The $\omega$ phase is very far from equilibrium at atmospheric pressure. The occurrence of a rapid transformation at low homologous temperatures ($T_m \approx$ 2128 K for Zr) indicated the $\omega$ was only weakly kinetically frustrated and that application of relatively small thermal driving forces was sufficient to overcome the transformation barrier. Yet, the asymptotic behavior, in which the transformation plateaus to another metastable equilibrium with retained $\omega$, suggests a more stable arresting mechanism. The question posed by the previous experiments can be phrased briefly as ``What is the source for this type of hysteresis and how does it simultaneously explain both the transient and asymptotic behavior of the transformation?'' 

\subsection{Prior Modeling and Simulation}
\label{sec:prior_model}
Nisoli et al. proposed a model which predicted the asymptotic stability of the $\omega$ phase under isothermal annealing, however this model did not address the initial rapid transformation rate during the initial stages of heating \cite{nisoli2016}. Nisoli et al. evolved the dislocation densities based on the Kocks and Mecking formulation \cite{kocks2003} that relates the dislocation evolution to shear evolution via a phenomenological model:
\begin{linenomath*}
\begin{equation}
\dfrac{d\rho}{d \gamma} = c_1\sqrt{\rho} - c_2\rho.
\end{equation} 
\end{linenomath*}

The source of shear was assumed to be a combination of  thermally activated plasticity and phase transformation (transformation shear). For the plastic shear contribution, Nisoli et al. takes the Orowan equation \cite{orowan1934} $\dot{\gamma}_p = \rho_m b v$, implying that plastic shear occurs by the movement of mobile dislocation populations present in the $\omega$ and that in turn contributes to dislocation elimination via the Kocks-Mecking relation listed above. Coupling between the dislocation evolution and the phase transformation was done in a similar manner to the modeling of martensitic transformations in TRIP-steels by Stringfellow et al. \cite{stringfellow1992}, in that the phase volume fraction $v_\omega$ is related to the shear $\gamma$ via: 
\begin{linenomath*}
\begin{equation}
\dot{v}_{\omega} \propto - v_\omega \dot{\gamma}.
\end{equation}
\end{linenomath*}
The reverse transformation rate and the dislocation evolution were described to be mutually coupled, in that occurrence of both processes lead to a net shear which in turn drives further dislocation evolution and phase reversion. One can think of this as a feedback loop in which the input and output continuously drive another. In the context of an asymptotic framework, this description is adequate in that the initial work required to kick-start the microstructural evolution has been done and does not require specification. While the model adequately captures the long-time behavior of the $\alpha/\omega$ microstructure, it does not address the initial transient phase of the transformation.

In a different study, Zong et al. performed a molecular dynamics simulation of the reverse transformation. The kinetics of the reverse transformation were found to follow a modified Kohlrausch--Williams--Watts equation, which is  sometimes used to describe ``glassy'' or frustrated kinetics \cite{takano1988}. Zong was also able to extract  a relaxation time constant which could be related to the thermal activation energy \cite{zong2014}. In said work, the authors noted that (a) the effective activation energies were higher in samples shocked to higher peak shock pressures (8 vs. 10.5 GPa); (b) the transformation can be modeled as a thermally activated process resulting from heterogenous nucleation from defects (ie. $\alpha/\omega$ interface, dislocations, grain boundaries, etc.). The molecular dynamics study of Zong suggested that the reverse transformation via interfacial growth of existing martensitic $\alpha$ laths was less likely to occur than the defect mediated heterogenous nucleation process at all temperatures, which was inconsistent with the ex-situ annealing experiments from Low et al. \cite{low2015}.  This apparent disagreement between molecular dynamics predictions and experimental observation indicates that a more detailed description of the defect states and their role in the reverse transformation is warranted. 

\section{Model for the Evolution of the Two-Phase $\alpha/\omega$ Microstructure} \label{section:numericalmodel}
In this work we present a temperature dependent model that couples the static recovery of dislocations with that of the $\omega\to\alpha$ transformation. Instead of coupling both processes via the evolution of shear, the dislocation density reduction is driven by the system's tendency to reduce the overall free energy. This reduction is subsequently coupled with the reverse transformation using an expression that describes the resistance to the reverse transformation in terms of the remaining dislocation content. 

\subsection{Dislocation Density Evolution Model}
At the beginning of the isothermal heating experiments, the post-shock defect state of the $\omega$ phase contains a high dislocation density arranged in a complex tangled dislocation network \cite{Song1994}. Under an annealing process, the removal of dislocation occurs by recombination or elimination reactions, which ultimately leads to the growth and maturation of these dislocation networks \cite{nes1995}. The proposed model for the evolution of the dislocation network is based on the framework of Nes \cite{nes1995} for the modeling of static recovery in iron, aluminum, and AlMg alloys. Nes considers the reduction of dislocation density through the growth and refinement of dislocation networks; which occurs through the processes of thermally activated glide, thermally activated cross slip, climb and solute drag as potential rate limiting mechanisms.

Following Nes, we assume these reactions have different rates depending on the underlying mechanism of defect annihilation. Given the high purity of the Zr samples we can neglect solute drag. We have also decided not to consider thermally activated cross slip, as we do not have a sufficient understanding of the likely deformation modes in $\omega$ and the relative mobilities of different dislocation populations especially in the as-shocked condition. Rather than introducing ad-hoc fitting parameters to account for the relative ease of the different mechanisms, we have decided to simplify the framework and only consider two rate limiting mechanisms:
\begin{enumerate}
	\item glide controlled: based on glide mobility of heavily jogged screw dislocations.
	\item climb controlled: based on the climb of edge dislocations.
\end{enumerate}

Our supposition is that during the early stage of the annealing process, sufficient energy is provided to activate network growth that is glide controlled. Upon exhaustion of defect populations that eliminate by means of glide controlled processes, the network growth then proceeds by removal of dislocations by a climb controlled process. Note that the rate limiting step for thermally activated glide is the motion of jogs in screw dislocation segments. Therefore, both glide and climb controlled reactions are fundamentally rate-limited by the climb of edge segments. 

In network growth, the average distance between dislocations, $r$, increases at a rate of 
\begin{linenomath*}
\begin{equation}
\dfrac{dr}{dt} = k_1 v
\label{eq:growthrate}
\end{equation}
\end{linenomath*}
Here, $k_1$ is a proportionality constant that, ideally, would be characterized experimentally. $v$ is the average velocity of dislocations
which depends on a driving force associated with a decrease in free energy by network growth: 
\begin{linenomath*}
\begin{equation}
F = \dfrac{k_2\\
	\mu b^2}{r}
\end{equation} 
\end{linenomath*}
where $k_2$ is a constant of order unity, $\mu$ is the bulk modulus, and $b$ is the burgers vector magnitude. 
 
\subsubsection{Glide Controlled Elimination}
\begin{figure}
	\centering
	\includegraphics[scale=0.08]{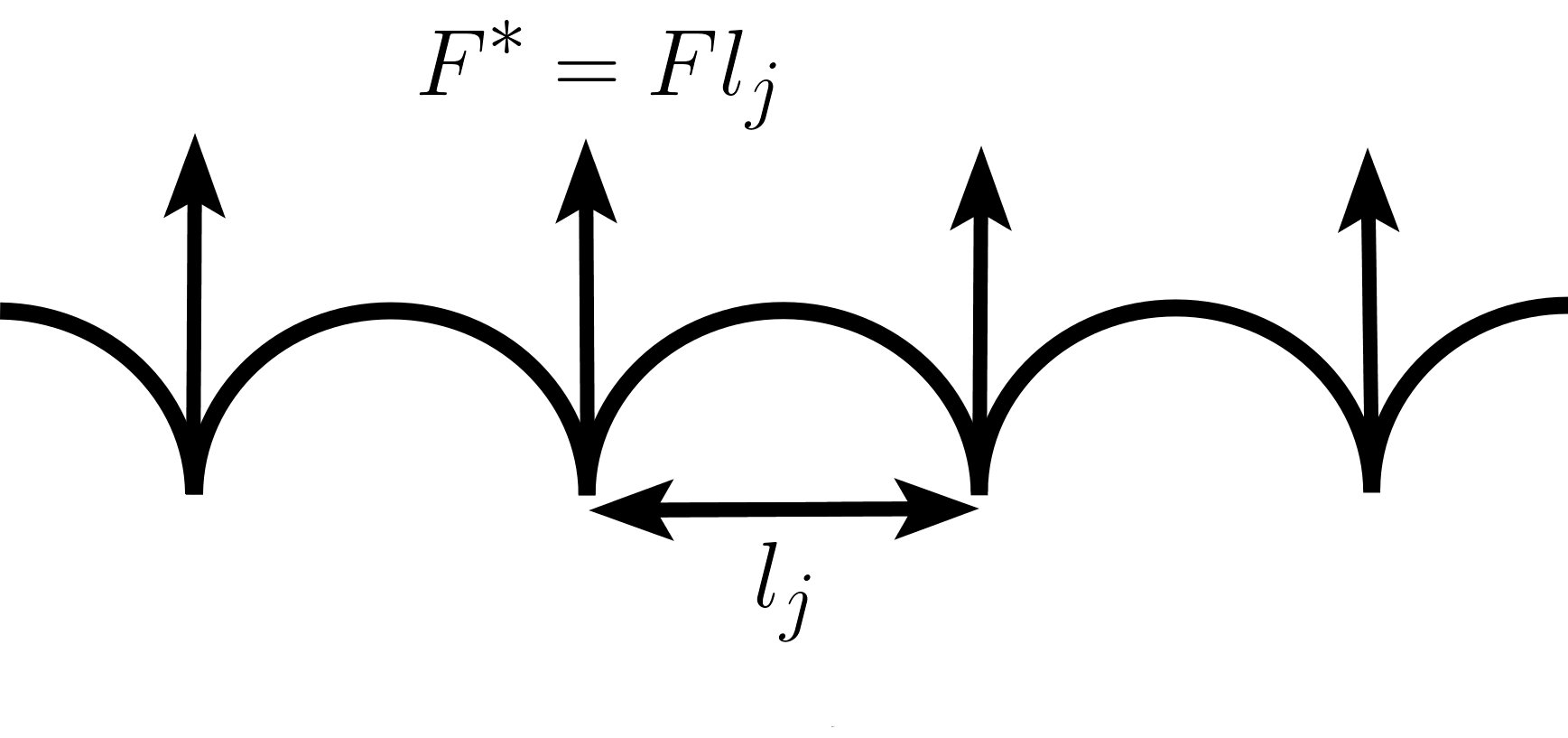}
	\caption{Bow-out of a jogged screw with average jog separation $l_j$. $F$ represents a force acting on the dislocation that is sourced from the decrease in free energy change associated with reduction in dislocation density, and $F^*$ represents a force enabling climb.}
	\label{fig:screw_jog}
\end{figure}

To derive an expression for the glide speed $v$, we consider that in a high purity metal, the rate controlling mechanism is expected to be the climb of jogs with edge character \cite{nes1995}. Consider an initially straight screw dislocation that contains jogs of alternating signs separated at a length of $l_j$ along the dislocation, as shown in figure \ref{fig:screw_jog}. Under the driving force $F$ a local vacancy equilibrium will be established at each jog 
\begin{linenomath*}
\begin{equation}
c=c_0 \exp{\pm(Fbl_j/rk_B\theta)}
\end{equation}
\end{linenomath*}
where $k_B$ is the Boltzmann constant, $\theta$ is the temperature. The $\pm$ indicates that jogs of opposite sign will respectively either emit or absorb vacancies during climb. The consequent presence of concentration gradients means that the rate of climb will be controlled by vacancy diffusion between jogs of opposite signs \cite{hirth1996}. In the case of bulk self diffusion at lower stresses, one can expect the dislocation mobility to be controlled by the slower moving vacancy absorbing jogs. The vacancy producing jogs also move while leaving vacancies behind, but would exert a weaker dragging force on the dislocation \cite{hirth1996}. Considering this, the glide speed can be expressed as: 
\begin{linenomath*}
\begin{equation}
v = \dfrac{4\pi k_3D_s}{b} \sinh \left (\dfrac{Fbl_j}{k_B\theta} \right ) \simeq \dfrac{4\pi k_3D_s}{b} \exp \left (\dfrac{Fbl_j}{k_B\theta} \right )
\end{equation}
\end{linenomath*}
where $D_s$ is the bulk self diffusivity and $k_3$ is a constant that needs to be determined experimentally. The $\sinh{}$ function has been replaced with the exp function on the assumption that $Fbl_j/k_B\theta > 1$. Combining the glide speed expression with equation (\ref{eq:growthrate}), the network growth rate that is climb controlled is given as: 
\begin{linenomath*}
\begin{equation}
\left( \dfrac{dr}{dt} \right)_{glide}  = \dfrac{4 \pi k_1 k_3 D_s}{b} \exp{ \left( \dfrac{Fb l_{j0} }{k_B\theta} \right) }
\label{eq:glidegrowth}
\end{equation}
\end{linenomath*}
$l_{j0}$ is the initial average jog spacing of the heavily jogged screw dislocation. For simplicity we have chosen to keep the average jog spacing constant for this model. The physical meaning of this assumption is that the lateral-jog-glide-mobility is low and the frequency of annihilation of opposite signed jogs is also low. This is a reasonable assumption in metals with low stacking faults or in solid solution strengthened alloys where glide is impeded by dislocation-solute interactions. Neither of these cases directly apply to  $\omega$-Zr as the SFE of $\omega \approx 479$ mJ/m$^2$ based on atomistic modeling \cite{kumar2017}. While this may be a less than satisfactory assumption, there is not a physically motivated way to estimate the lateral-jog-glide-mobility. Rather than introducing an additional fitting parameter in the model to control this behavior, we felt that the assumption of a constant jog spacing was a better choice based on the limitations on the current state of understanding of $\omega$ phase defects. The jog spacing is related to the average dislocation separation by: 
\begin{linenomath*}
\begin{equation}
l_{j0} = \kappa r_0
\label{eq:kappa2}
\end{equation} 
\end{linenomath*}
where $\kappa$ is a proportionality constant, and $r_0$ is the initial average separation between dislocations. $\kappa$ can be interpreted as a proxy of the extent of deformation experienced during the shock -- a lower value of $\kappa$ indicates a higher extent of prior deformation. A higher proportion of mobility limiting jogs indicate relatively more interactions during plasticity of the shock deformation. We would expect $\kappa$ to be lower in shocked samples than in quasi-static deformed samples. 

\subsubsection{Climb Controlled Elimination}
In the case of the annihilation of two edge dislocations, the movements of the dislocations prior to annihilation can be conceptually decomposed into that of a pure glide of length $l_g$ followed by a pure climb of length $l_c$. By treating the overall glide/climb movement as a summation of elementary activations (each of climb step $b$), the activation length can be written as $l_a = \xi b$, where $\xi = 1+l_g/l_c$. Again assuming a vacancy diffusion controlled climb for bulk diffusion, the average dislocation speed for climb control can be expressed as \cite{hirth1996}: 
\begin{linenomath*}
\begin{equation}
v = \dfrac{2 \pi \xi D_s }{b \ln{\left ( \dfrac{2r}{b} \right)}} \left [ \exp{\left ( \dfrac{Fb^2}{k_B \theta}\right )} - 1\right ]
\end{equation}
\end{linenomath*}
which upon combination with equation \ref{eq:growthrate} results in the following network growth rate:
\begin{linenomath*}
\begin{align}
\left (\dfrac{dr}{dt}\right )_{climb} &= \dfrac{2 \pi k_1 \xi D_s }{b \ln{\left ( \dfrac{2r}{b} \right)}} \left [ \exp{\left ( \dfrac{Fb^2}{k_B \theta}\right )} - 1\right ]\\
&= \dfrac{2 \pi \tilde{k} D_s }{b \ln{\left ( \dfrac{2r}{b} \right)}} \left [ \exp{\left ( \dfrac{Fb^2}{k_B \theta}\right )} - 1\right ]
\label{eq:climbgrowth}
\end{align} 
\end{linenomath*}
It is worth mentioning that the determination reasonable values of $\xi$ is non-trivial. Again, to reduce the number of fitting parameters in the absence of experimental values $\xi$ and $k_1$ have been combined into $\tilde{k}$. 

\subsubsection{Dislocation Evolution}
The total growth rate of the dislocation networks is then just a sum of the aforementioned contributions: 
\begin{linenomath*}
\begin{equation}
\dfrac{dr}{dt} = \left (\dfrac{dr}{dt}\right )_{glide} + \left (\dfrac{dr}{dt}\right )_{climb}
\end{equation}
\end{linenomath*}
Approximating the dislocation density as $\rho \approx (1/r^2)$, we are able to  evolve the dislocation density with respect to time via: 
\begin{linenomath*}
\begin{equation}
\dfrac{d\rho}{dt} = -2\rho^{ \frac{3}{2}} \dfrac{dr}{dt}
\end{equation} 
\end{linenomath*}
from which the dislocation density can be used to calculate the microstrains via:
\begin{linenomath*}
\begin{equation}
\label{eq:dd1}
	\varepsilon = \sqrt{\dfrac{\rho}{\rho_0}} \varepsilon_{0}
\end{equation}
\end{linenomath*}
where $\rho_0$ and $\varepsilon_0$ are the initial dislocation density and experimentally measured microstrain respectively.

\subsection{Phase Transformation Model}
The rate at which the $\omega$ phase volume fraction $f_\omega$ varies with time is given as a double exponential rate equation:  
\begin{linenomath*}
\begin{equation}
\dot{f}_\omega = \dot{f}_0 f_\omega \exp{\left (  \dfrac{\alpha_1\Delta G}{R \theta}\right )} \exp{\left ( \dfrac{\beta_1 B(\rho)}{k_B \theta}\right )}
\label{eq:transformation_model}
\end{equation}
\end{linenomath*}
where $\dot{f}_0$ is a reference transformation rate and $R$ is the gas constant. Conceptually, one can think of the first exponential term being that of the driving force for the $\omega \rightarrow \alpha$ transformation due to the Gibbs free energy differential $(\Delta G)$  at the current temperature and pressure. This driving term is assumed to be independent of the current dislocation state of the $\omega$ phase. $\alpha_1$ is a unitless parameter which increases non-linearly with temperature and can be interpreted as an indicator for the specification of the $\omega\to\alpha$ transformation mechanism (ie. growth of existing $\alpha$ laths vs nucleation and growth of new $\alpha$ plates). 
$\Delta G$ is evaluated using an equation of state (EOS) for Zr detailed by Greeff \cite{greeff}. The EOS for Zr provides the Helmholtz free energy, $H^{eos}$, via: 
\begin{linenomath*}\begin{equation}
H^{eos}(V,\theta) = \phi_0(V) + H^{ion}(V,\theta) + H^{el}(V,\theta)
\end{equation}\end{linenomath*} 
where $\phi_0$ is the static lattice energy, $H^{ion}$ is the ion motion free energy and $H^{el}$ is the electronic excitation free energy. According to Greeff, $\phi_0$ dominates the pressure or volumetric contribution to the Helmholtz free energy expression, while $H^{ion}$ has the strongest temperature dependence. For the analytical expressions and calibrated parameters for these individual contributions, and the reader is referred to the work of Greeff \cite{greeff2004modeling}. Consequently, the Gibbs free energy for each phase can be calculated as a function of specific-volume and temperature via:
\begin{linenomath*}\begin{equation}
	G = H^{eos} + Pv
\end{equation}\end{linenomath*}
where $P$ is the pressure and $v$ is the specific volume of each corresponding phase. 

The second exponential term in Eq. \ref{eq:transformation_model} describes the barrier energy, due to the defect state of the microstructure, that resists the $\omega\to\alpha$ transformation under heating. The barrier energy is related to the energy required to mobilize the arresting dislocations, subsequently allowing elimination reactions to occur. Not all dislocations will have the same mobility or be equivalent in their propensity to arrest the transformation. Here we make the assumption that effectiveness of a dislocation at resisting the transformation is directly related to the energy requirement for mobilizing that defect. Experimentally, Low et al. observed a large decrease in the $\omega$ phase dislocation density before any significant transformation occurred. Here, we postulate that the highly mobile defects exert little to no arresting influence and due to their high mobility are removed early during the recovery process. By this logic, the barrier energy or the resistance to the $\omega\to\alpha$ transformation should increase as the dislocation density is reduced as the remaining dislocation population are less mobile and more effective at retarding the transformation. The barrier energy is evaluated as follows:
\begin{linenomath*}
\begin{subequations}
	\begin{align}
	B(\rho) &= \dfrac{E_{climb} + E_{glide}}{\rho} \label{eq:barrierenergy_total}\\
	E_{climb} &= Fb^2\\
	E_{glide} &= Fbl_j
	\end{align}
	\label{eq:barrierenergy}
\end{subequations} 
\end{linenomath*}
The expressions for the individual energy contributions are from the exponential arguments of the individual dislocation network growth rate equations (eq. (\ref{eq:glidegrowth}) and (\ref{eq:climbgrowth})). Note that the energy terms in equation \ref{eq:barrierenergy} are not constant, as the driving force $F$ varies inversely proportional to the average dislocation separation. Taking into account that the barrier energy $B$ represents an averaged activation energy for mobilization of \textit{individual} dislocations, the $\beta_1$ parameter can be thought to be a scaling of this energy to represent a reference defect population providing this resistive energy, and is on the order of $10^{-1} \rho_{ref} \approx 10^{14}$.

\subsection{Model Parameters} \label{subsection:modelsummary}
Table \ref{table:dislocation_parameters} is a summary of the parameters used for the dislocation density evolution model. The parameters listed here are calibrated to the fitting of the annealing of 8 GPa samples.

\begin{table*}[]
	\centering
	\caption{Summary of material parameters for the dislocation model}
	\begin{tabular}{@{}llll@{}}
		\toprule
		Parameter           & Value         & Units    & Description                                              \\ \midrule
		$b$                 & 5.034656e-10  & m        & burgers magnitude                                        \\
		$\mu$               & 35366.667e+06 & Pa       & bulk modulus                                             \\
		$k_2$          & 1.55          & -        & dislocation energy fudge factor                          \\
		$\kappa$          & 0.023         & -        & jog/dislocation separation ratio                         \\
		$k_1$               & 1.0e-06       & -        & exponential pre-factor (glide + climb)                   \\
		$k_3$               & 5.0e-05       & -        & exponential pre-factor (glide)                           \\
		$\tilde{k}$               & 1.0e-06           & -        & exponential pre-factor (climb)                           \\
		$\rho_{ref}$        & 1.0e+15       & m$^{-2}$ & reference dislocation density  \\
		$\varepsilon_{ref}$ & 0.0126        & -        & reference microstrain  \\ 
		\bottomrule                        
	\end{tabular}
	\label{table:dislocation_parameters}
\end{table*}

The value of the exponential pre-factors ($k_1, k_3, \tilde{k}$) from equations (\ref{eq:glidegrowth})and (\ref{eq:climbgrowth}) are not readily obtainable from experimentation. Even though they have a clear physical significance there is no way to directly estimate the parameters or constrain them with physically motivated bounds. Since these terms affect the outcome of the calculation of the network growth rate in a similar manner to the self diffusivity $D_s$ by acting as an exponential pre-factor, we can consider the product of the these parameters in Eq. \ref{eq:glidegrowth} to be an \emph{effective diffusivity} and verify that it is consistent with the limited experimental observations. Zr is known to have vanishingly low self diffusivities at low homologous temperatures ($\approx10^{-23}$ at 800 K) \cite{Horváth1984206}. However, it has been reported that precipitation of the $\omega$ phase has been correlated with an anamolous increase in diffusivity \cite{sanchez1978}. For this work we have chosen, somewhat arbitrarily, for the diffusivity at 443 K to be $\approx10^{-22}$.  By keeping $k_1$, $k_3$, $\tilde{k}$ as temperature independent parameters, we can extract the trend in diffusivities with increasing temperature.  

In previous work, Brown et al. assumed the dislocation density as being proportional to the microstrains reported from the XRD analysis \cite{Brown2014383, low2015}. For the $\alpha$ phase, this can be made quantitative through the application of line profile analysis. Line profile analysis requires detailed knowledge of the available slip systems and likely dislocations present \cite{balogh2009twinning}; data which is not currently available for $\omega$. For the $\omega$ phase, a problem arises in that we have quantitative values of the microstrain but no direct measurement of the dislocation density within the $\omega$ phase through conventional line profile analysis \cite{Brown2014383}. This means that the initial $\omega$ phase dislocation density needs to be estimated. We chose the 443 K sample shocked to 8 GPa as a baseline reference sample. Based on a measured $\alpha$ phase dislocation density of  $\rho^{\alpha}_0 \approx 9\times 10^{14}$ m$^{-2}$ \cite{Brown2014383}, we chose to set the initial $\omega$ dislocation density to be $ \approx 10^{15}$ m$^{-2}$ for this sample. By taking the initial values of the microstrain and dislocation density of the 443 K sample shocked to 8 GPa as reference microstrain $\varepsilon_{ref}$ and reference dislocation density $\rho_{ref}$, respectively, the initial dislocation densities in the other samples can be calculated as:  
\begin{linenomath*}\begin{equation}
	\rho_0 = \left (\dfrac{\varepsilon_0}{\varepsilon_{ref}}\right )^2 \rho_{ref}
	\label{eq:rh0}
\end{equation}\end{linenomath*}

In table \ref{table:transformation_parameters}, a list of parameters related to the transformation model is listed. 
\begin{table*}[]
	\centering
	\caption{Parameters used for the reverse transformation rate, equation (\ref{eq:transformation_model})}
	\begin{tabular}{@{}llll@{}}
		\toprule
		Parameter   & Value   & Units  & Description                                \\ \midrule
		$\dot{f}_0$ & -0.5    & s$^{-1}$ & reference transformation rate              \\
		$\beta_1$   & 1.9e+14 & m$^{-2}$ & effective resisting dislocation population \\
		\bottomrule
	\end{tabular}
	\label{table:transformation_parameters}
\end{table*}

\section{Results} \label{sec:results}
\subsection{Calibration to Isothermal Annealing Experiments} \label{subsection:isothermal}
\begin{figure*}
	\centering
	
	\subfloat[$\omega$ phase microstrains]{\includegraphics[scale=0.21]{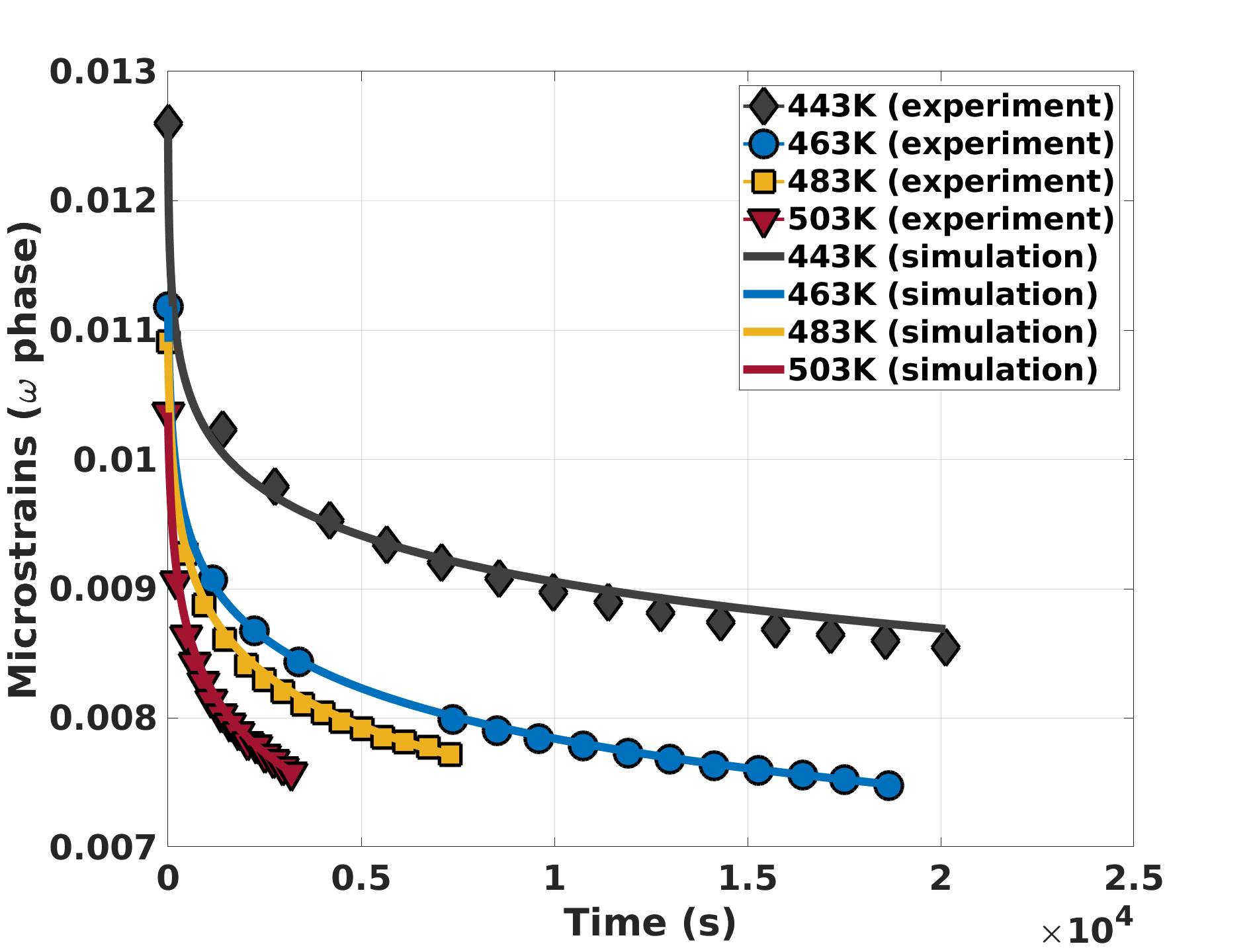}\label{fig:8GPa_omegaStrain}}
	\subfloat[$\omega$ phase volume fractions]{\includegraphics[scale=0.21]{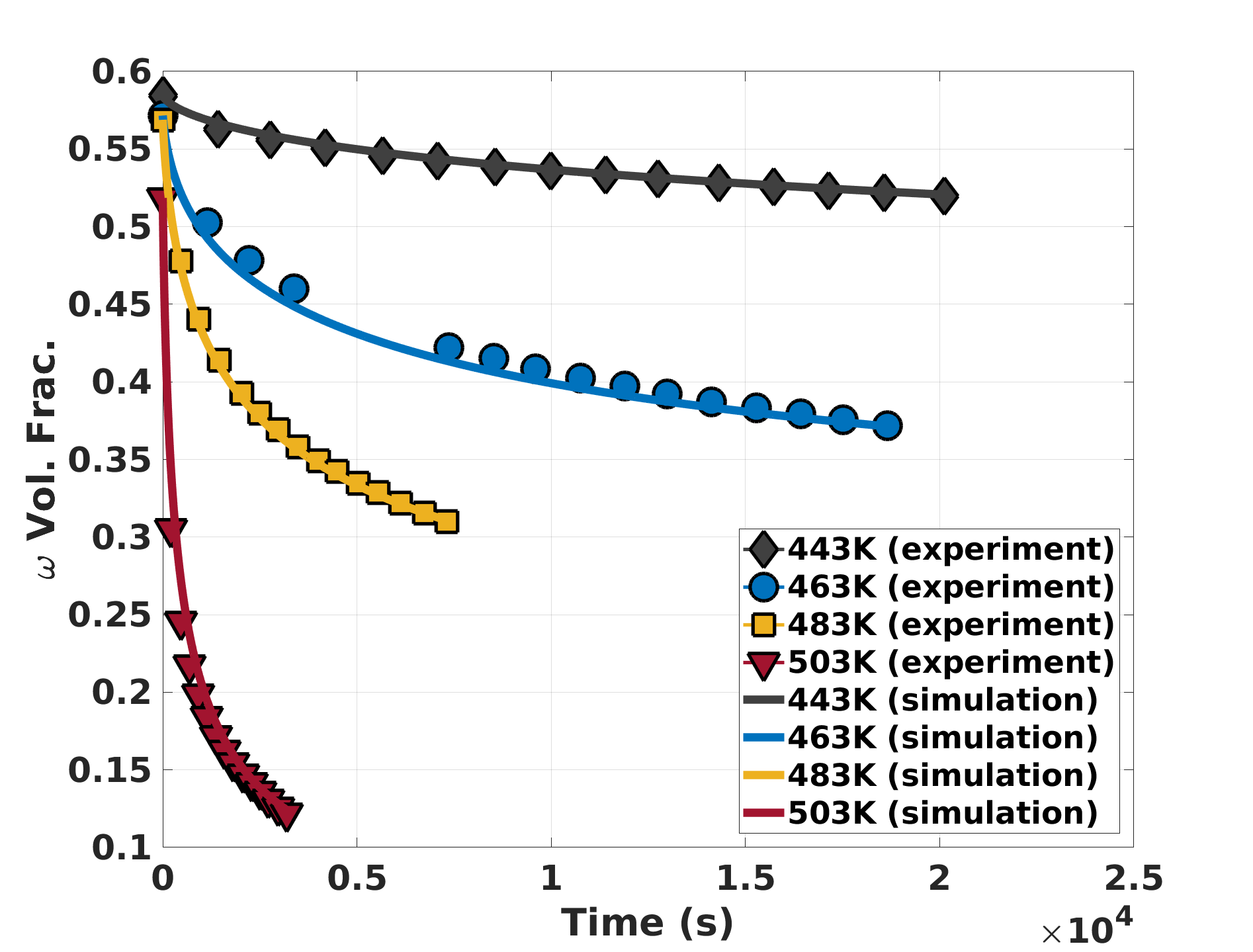}\label{fig:8GPa_omegaFrac}}

	\caption{Simulation fits for annealing of the Zr sample that was shocked to 8 GPa over time. The data markers have been reduced for purpose of clarity and the solid lines indicate simulation results.}
	\label{fig:8GPa_simulation}
\end{figure*}

Figure \ref{fig:8GPa_simulation} illustrates model fits to the isothermal annealing of the 8 GPa sample. Using the parameters provided in section \ref{section:numericalmodel} as a baseline, the remaining parameters that remain unknown are the self diffusivities of the $\omega$ phase that influences the dislocation removal rate and the $\alpha_1$ parameter of the transformation rate equation (equation \ref{eq:transformation_model}). Due to any lack of experimental reports for these parameters, the temperature dependent trends of these two parameters are treated as model outcomes and potential validation points should the diffusivities be measured or calculated in the future. The extracted values for these two parameters are provided in table \ref{table:fitted_parameters}. For the microstrains illustrated in figure \ref{fig:8GPa_omegaStrain}, the fits for the higher temperatures showed good correspondence with the samples annealed at higher temperatures (463-503 K) but showed a mild deviation in the curvature at 443K. The fits for the $\omega$ were also satisfactory and able to capture the initial rapid transformation and asymptotic behavior. 

\begin{table*}[]
	\centering
	\caption{Calibrated parameters for the $\omega$ phase self diffusivities, $D_s$ and the $\alpha_1$ parameter for the reverse transformation rate equation at multiple temperatures based on fitting to the isothermal annealing experiments.}
	\begin{tabular}{lllll}
		\toprule
		Parameter  & 443 K               & 463 K               & 483 K               & 503 K               \\ \cline{1-5} \\
		$D_s$ (m$^2$/s)      & $4.0\times10^{-22}$ & $7.0\times10^{-21}$ & $2.0\times10^{-20}$ & $9.0\times10^{-20}$ \\
		$\alpha_1$ (unitless) & 0.1                 & 5.9                 & 7.3                 & 11.3       \\
		\bottomrule        
	\end{tabular}
	
	\label{table:fitted_parameters}
\end{table*}

\begin{figure*}
	\centering
	
	\subfloat[$\omega$ phase microstrains]{\includegraphics[scale=0.21]{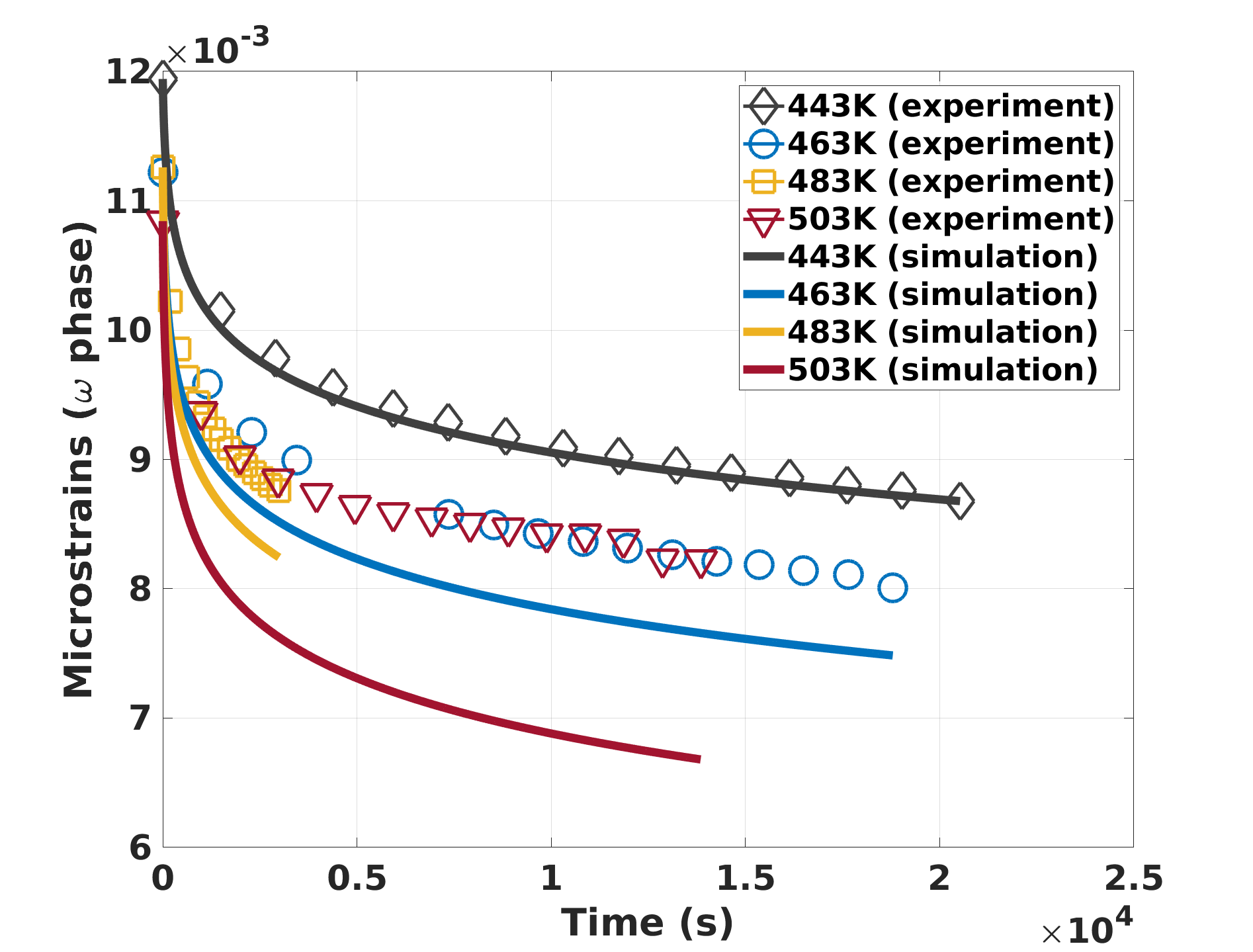}\label{fig:10GPa_omegaStrain}} 
	\subfloat[$\omega$ phase volume fractions]{\includegraphics[scale=0.21]{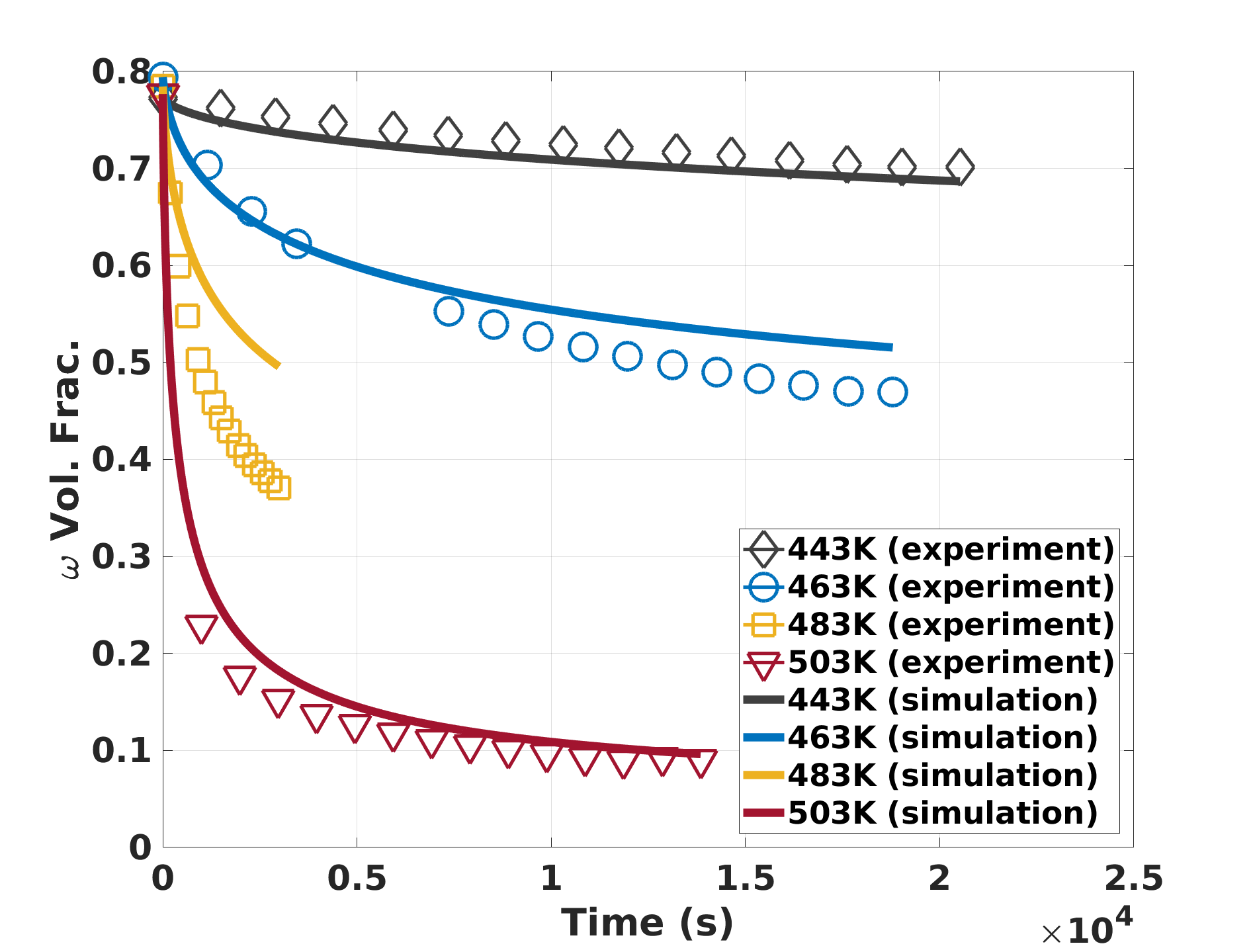}\label{fig:10GPa_omegaFrac}}
	
	\caption{Simulation fits for annealing of the Zr sample that was shocked to 10.5 GPa over time. The simulation results are based on parameters that were calibrated for the annealing of 8 GPa sample to test robustness of the model given a sample shocked to a different pressure.}
	\label{fig:10GPa_simulation}
\end{figure*}

The results of exercising the model on the 10.5 GPa shock samples is shown in figure \ref{fig:10GPa_simulation}. The parameters calibrated to the 8 GPa samples were retained, with only the initial condition (e.g. volume fractions and $\omega$ dislocation density) varied. As can be seen from the figure, the predictions of the $\omega$ phase fractions are reasonable. However, the model over-predicts the reduction of microstrain for all temperatures except 443K. Low et al. observed that the evolution of microstrain (and by extension dislocation density) of the 10.5 GPa samples was fundamentally different than the 8 GPa samples, leading to a speculation that the defect populations of the two samples were qualitatively different likely due to more extensive $\omega$ phase plasticity during the shock event \cite{low2015}. The model predictions are consistent with this hypothesis and suggest that within the 10.5 GPa samples there is a population of defects that are not readily annealed out at these low homologous temperatures but is also not actively resisting the $\omega \rightarrow \alpha$ transformation. This point is discussed further in section \ref{sec:discussion}.


\subsection{Effect of Jog Spacing}
\label{sec:jog}
\begin{figure}
	\centering
	\includegraphics[scale=0.25]{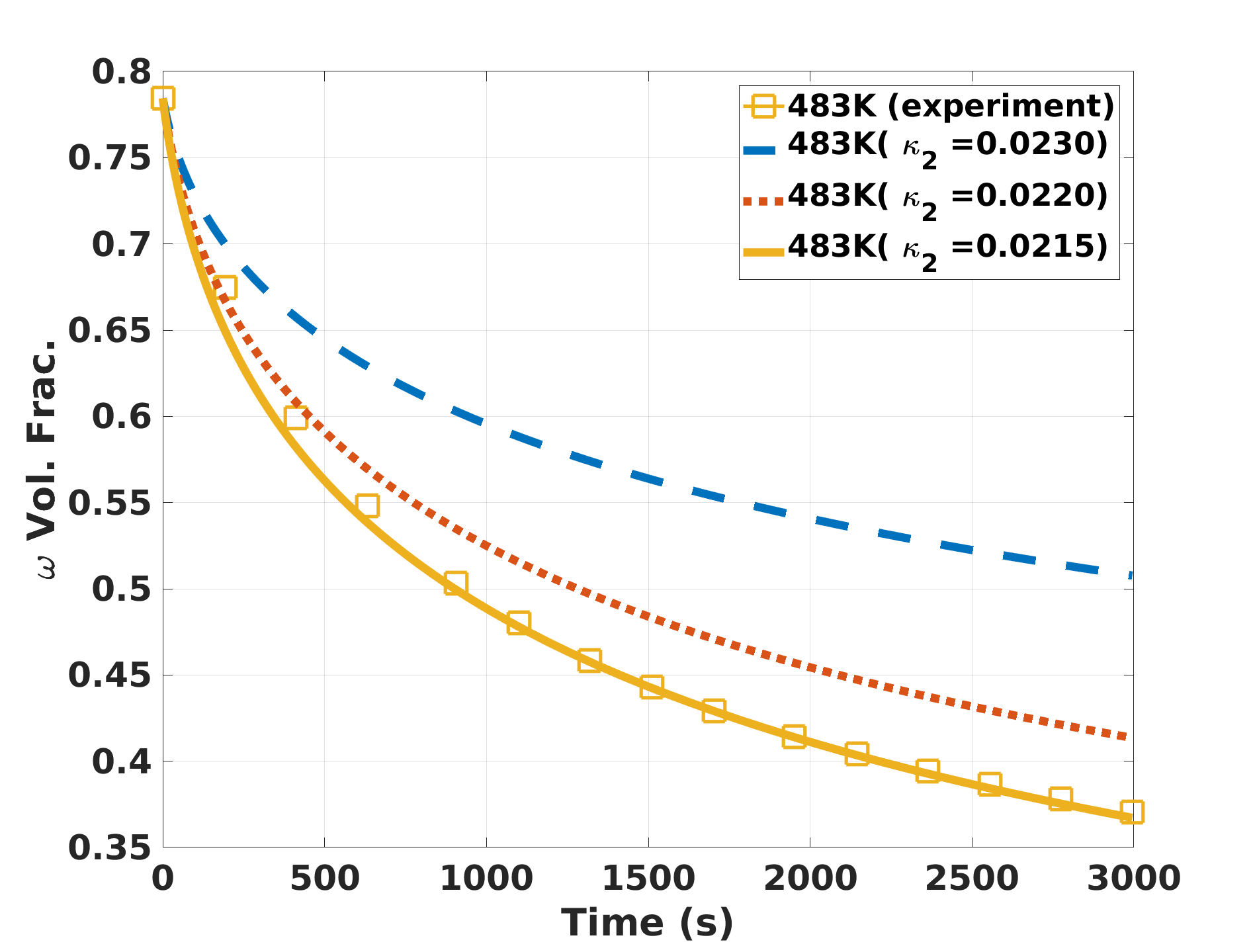}
	\caption{Comparing the effect of varying the $\kappa$ parameter from the glide controlled growth rate equation on the $\omega$ phase fraction evolution}
	\label{fig:kappa}
\end{figure}

The poor predictions for the $\omega$ volume fraction for the 483K 10.5 GPa sample (see figure \ref{fig:10GPa_omegaFrac}) can be sensibly and simplistically remedied by a small change in the value of $\kappa$ from 0.023 to 0.0215 as shown in figure \ref{fig:kappa}. The variation of $\kappa$ has a significant effect on the initial transient behavior of the $\omega$ volume fraction. As $\kappa$ decreases, the initial transformation rate increases. From a physical understanding, the adjustment of the $\kappa$ parameter is equivalent to varying the jog spacing on screw dislocations, which concerns the dislocation population that eliminates by glide controlled mechanisms. In a more deformed sample, one would expect a higher frequency of jogs along its dislocations. Again, this is consistent with the notion that the 10.5 GPa samples have undergone more severe plastic than the 8 GPa samples. As the degree of deformation is increased, the heterogeneity is also expected to increase. The in-situ annealing samples were small 3mm disks cut from the large impact specimen \cite{low2015}. It is not surprising to find spatial variation in defect state in the shocked samples and that the 8 GPa samples are more homogeneous than the 10.5 GPa samples. From this analysis it would seem that the jog spacing, $\kappa$, could serve as a good indicator of the plastic history of the samples.

\subsection{Validation: Linear Temperature Ramp}
As a validation of the framework, the model was exercised on the constant temperature ramp data extracted from Brown et al. \cite{Brown2014383}. As described above all model parameters except the diffusivities and $\alpha_1$ were considered to be temperature independent. In order to match/simulate the heating from room temperature to 620 K, the values for the diffusivities and $\alpha_1$ had to be extrapolated beyond the range given in Table \ref{table:fitted_parameters} to higher and lower temperatures. The self diffusivities were fit to a two term exponential equation. For the $\alpha_1$ parameter, a piecewise linear interpolation has been applied for the temperature range of 443-503 K. For temperatures in the range ($503 < \theta \le 620$ K) , this fit was continued by linear extrapolation to the point $(\theta, \alpha_1) = (620, 24)$. For temperatures less than 443 K, a linear extrapolation to the origin was chosen. Remember that the $\alpha_1$ parameter acts as a modifier to the bulk Gibbs free energy based to account for potentially different transformation mechanisms over different temperature ranges. 

\begin{figure*}
	\centering
	\subfloat[Self diffusivity]{\includegraphics[scale=0.21]{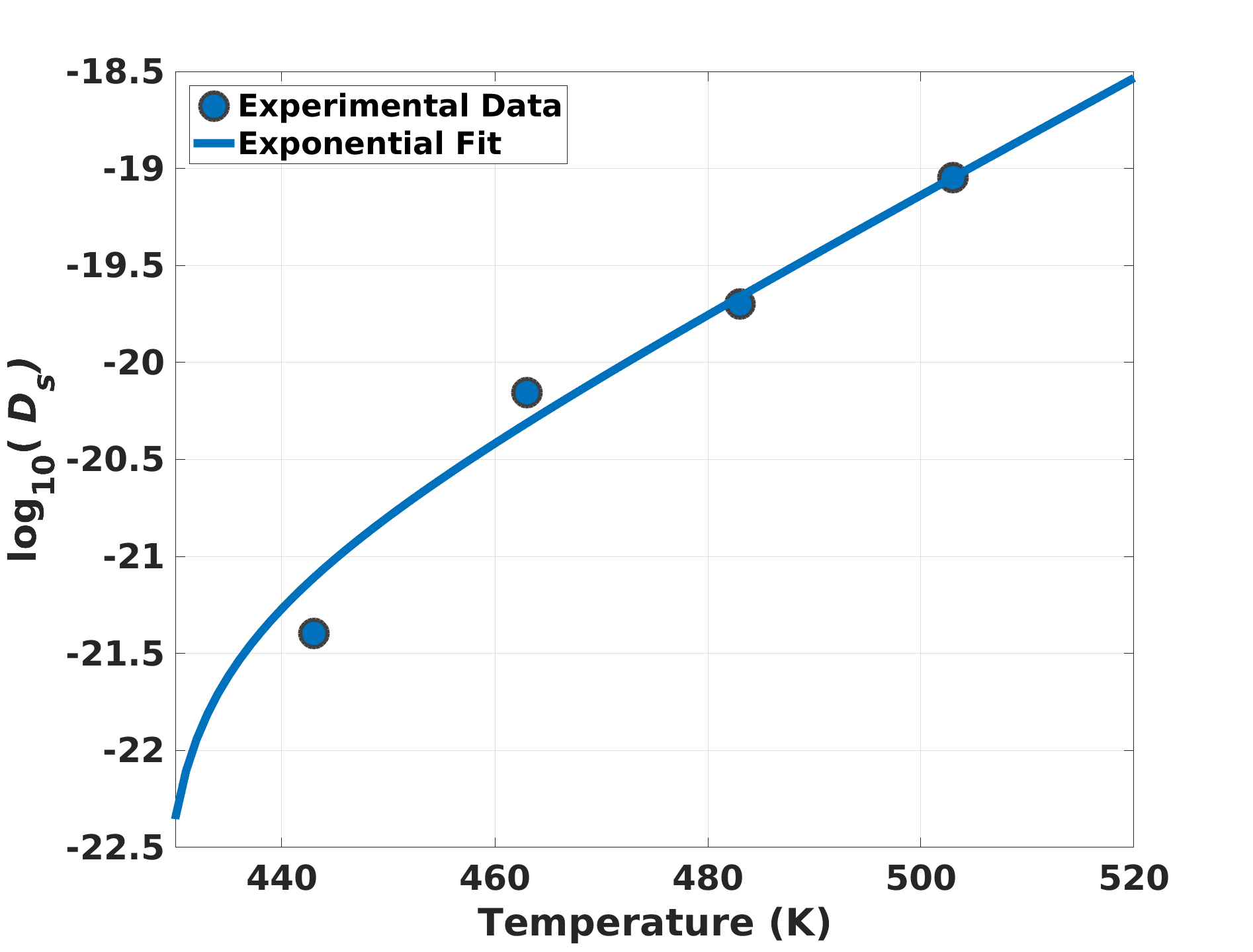}\label{fig:diffusivity}}
	\subfloat[$\alpha_1$ parameter]{\includegraphics[scale=0.21]{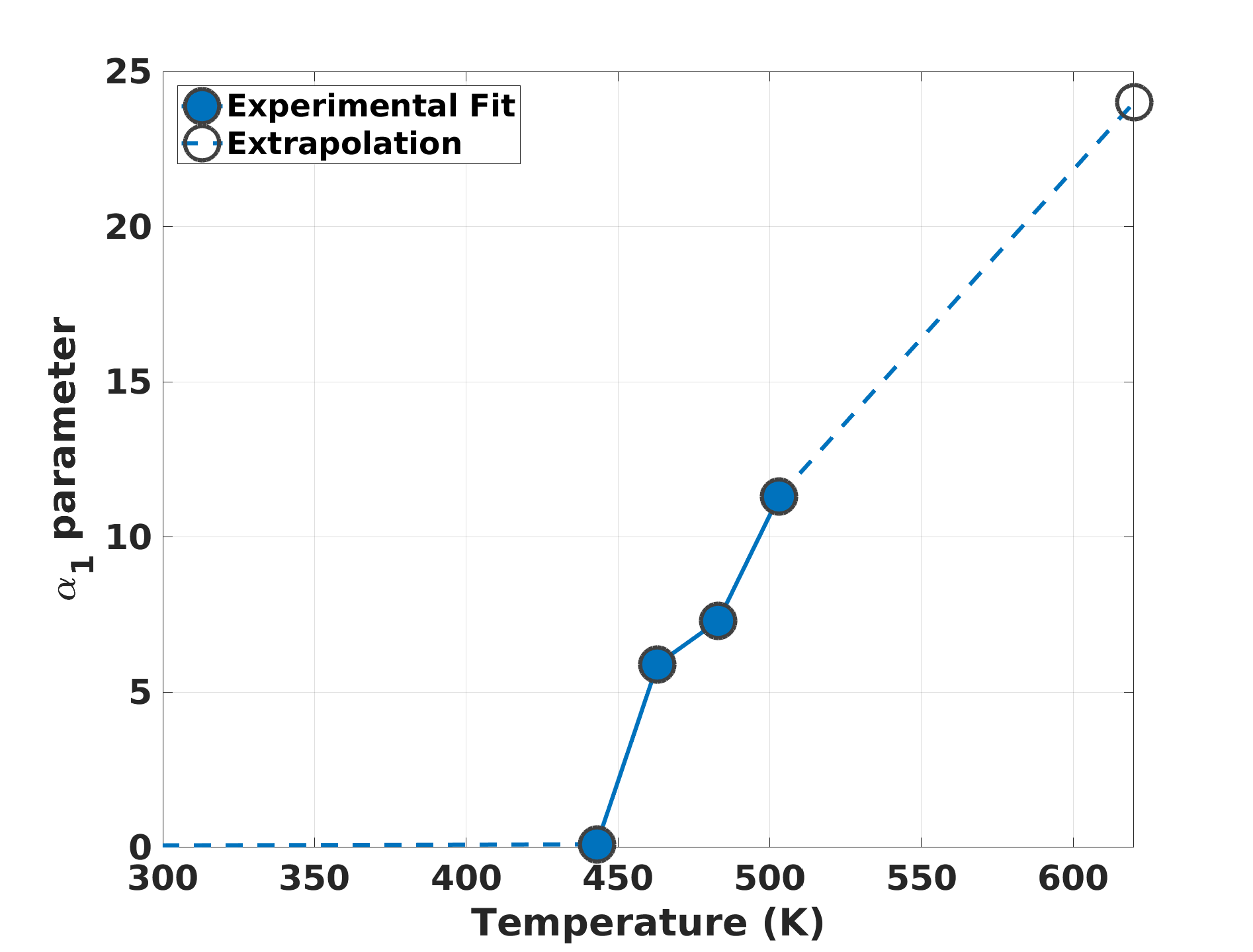}\label{fig:alpha1}}
	
	\caption{Calibrated temperature dependent data points from the model and their corresponding numerical fits for the (a) $\omega$ phase self diffusivity and (b) the $\alpha_1$ parameter. A piecewise linear fit is applied for the $\alpha_1$ parameter. }
	\label{fig:temperature_parameters}
	
\end{figure*}
\begin{figure}
	\centering
	\includegraphics[scale=0.25]{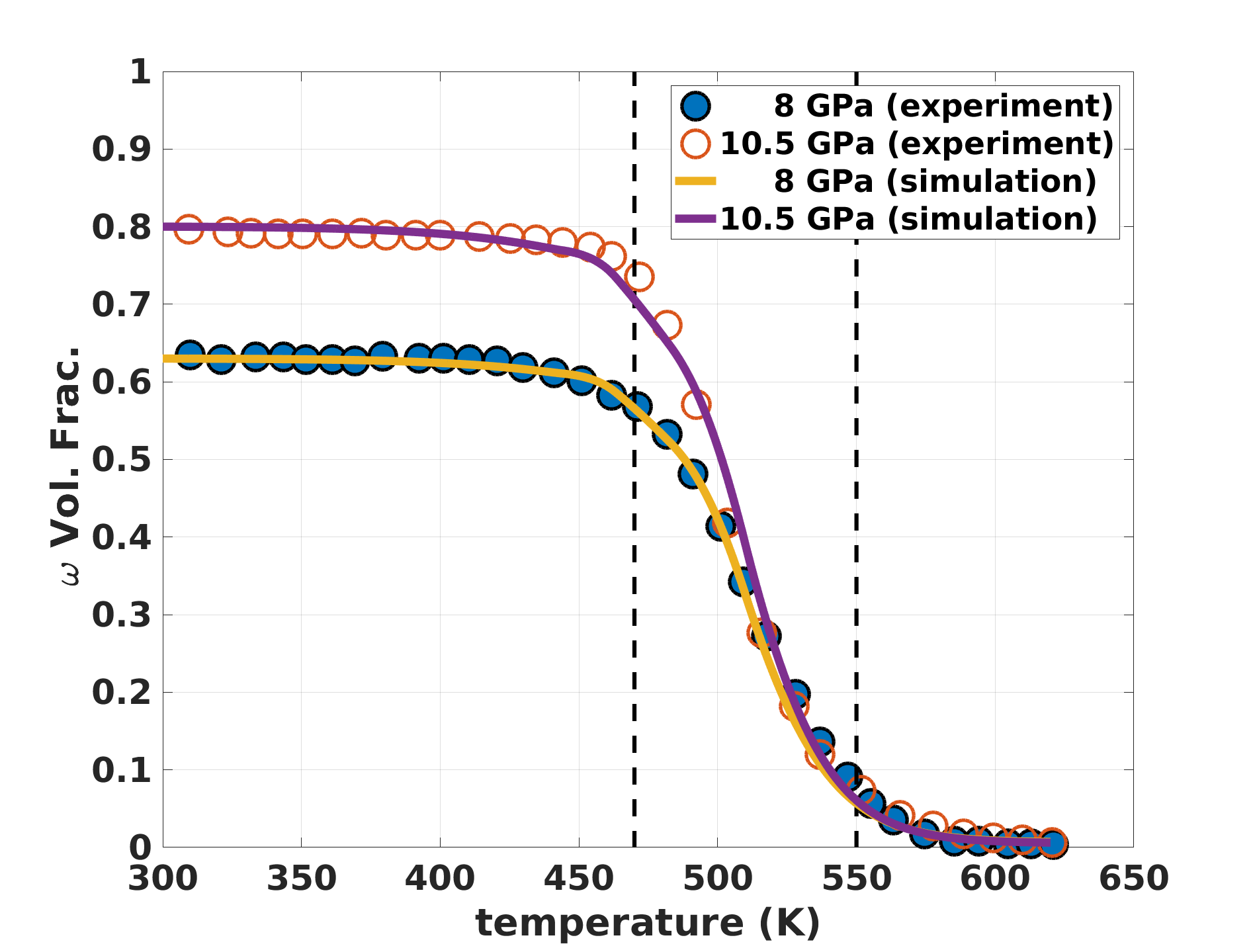}
	\caption{Exerimental data and simulated fits of the $\omega$ phase fraction under a constant heating rate of 0.05 K/s from 300 to 620 K. Experimental data is sourced from Brown et al. \cite{Brown2014383}.}
	\label{fig:constantRamp}
\end{figure}

The model predictions vs. the Brown et al. data is displayed in figure \ref{fig:constantRamp}. The figure visually confirms an excellent consistency between the model and the experimental values. With the exception of $\kappa$, all other model parameters remained fixed at the values calibrated for the 8GPa isothermal annealing case. $\kappa=0.217$ and $\kappa=0.210$ was used  for the 8 and 10.5 GPa constant heating ramp samples respectively. The changes represent a difference of approximately 6\% to our the isothermal 8 GPa values. Reiterating that the $\kappa$ value represents the jog spacing on screw dislocations, a 6\% difference in value is quite reasonable due to the annealing samples being obtained from different regions of the larger shock loaded Zr plate (Also the annealing samples were removed at different times, more than one year separated the sample prep for the Brown at al. and Low et al. studies). Differences of a few percent in the initial volume fraction of $\omega$ phase also points to the samples from both experiments seeing slightly different deformation conditions as well.

\section{Discussion} \label{sec:discussion}
\subsection{Glide vs. Climb Controlled Dislocation Removal} \label{subsection:discussion_ddremoval}
One of the key features of the current model is the concept of elimination of dislocation populations by two modes, more specifically the glide and climb controlled removal processes. Nisoli et al. had previously modeled the reverse transformation during annealing in the limit of long times \cite{nisoli2016} and an overview of that work was presented in section \ref{sec:prior_model}. Nisoli et al. starts with a Kocks Mecking formulation in which the dislocation density is evolved in relation to shear, and considers two sources for shear: i) thermally activated dislocation motion and ii) transformation shear. Nisoli further makes the assumption, based on experimentally observed transformation rates, that the transformation rate decreases to zero as time goes to infinity and all the shear can be assumed to come from thermally activation of dislocations at asymptotic times. By considering a Taylor expansion of the Kock Mecking equation around the point $\rho(t)=\rho(\infty)$,  Nisoli et al. eventually derives an expression for the dislocation evolution at long times, as shown: 
\begin{linenomath*}\begin{equation}
\label{eq:nisoli}
\rho = \dfrac{\rho_0}{(1+t/\tau)^{1/\psi}}
\end{equation}\end{linenomath*}
where is a $\tau$ is a critical timescale which scales with the temperature dependent mobility of the dislocation population. This equation provides a very crucial and insightful link, as it is of the same form that Nes derives for climb controlled dislocation elimination in annealing processes \cite{nes1995}. Nes also notes that the climb controlled elimination processes dominates when glide contribution to the dislocation removal is exhausted, therefore providing a key to the transient vs asymptiotic behavior of the dislocation removal trends. 

\begin{figure*}
	\centering
	
	\subfloat[443 K ]{\includegraphics[scale=0.21]{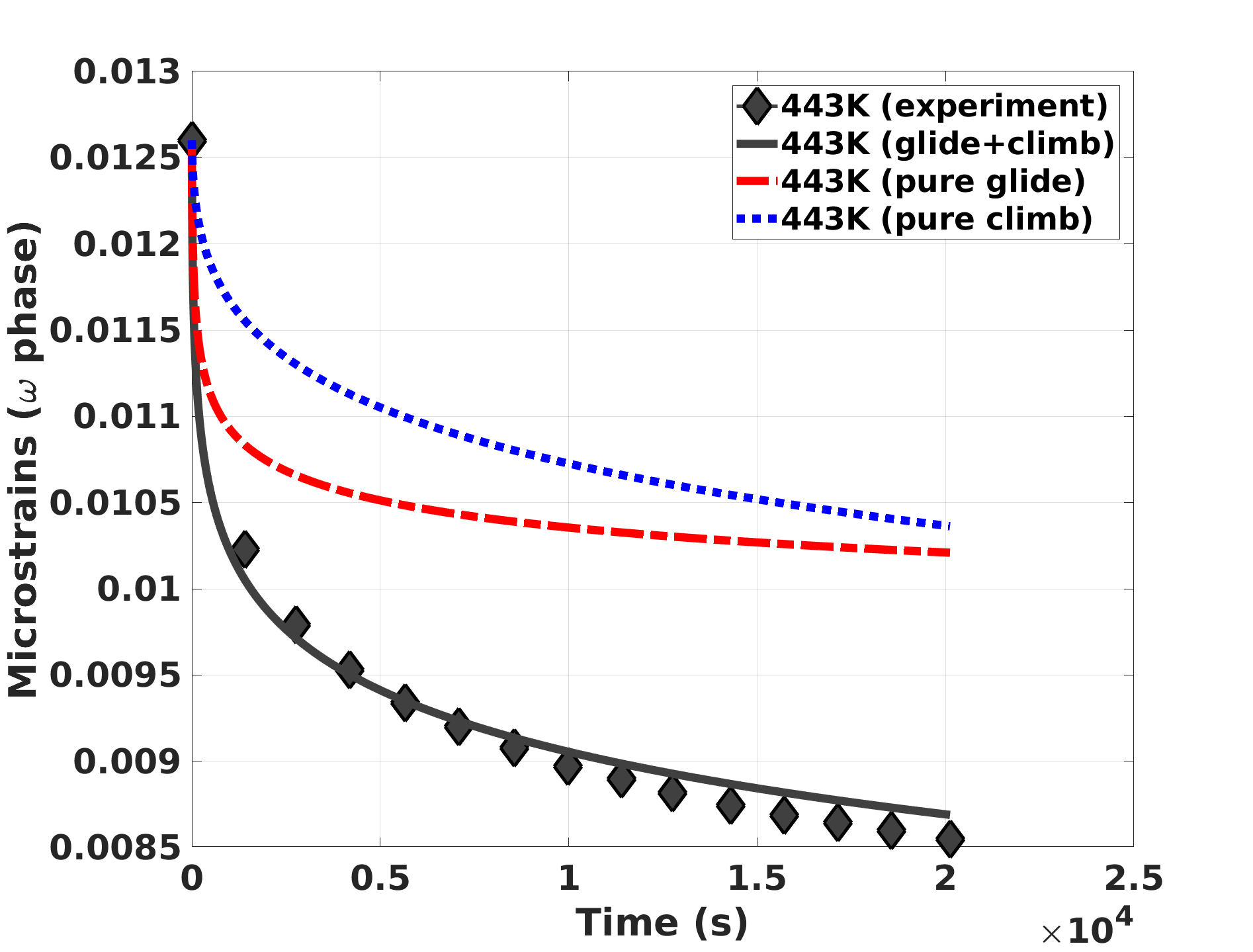}\label{fig:170C_glideclimb}}
	\subfloat[503 K]{\includegraphics[scale=0.21]{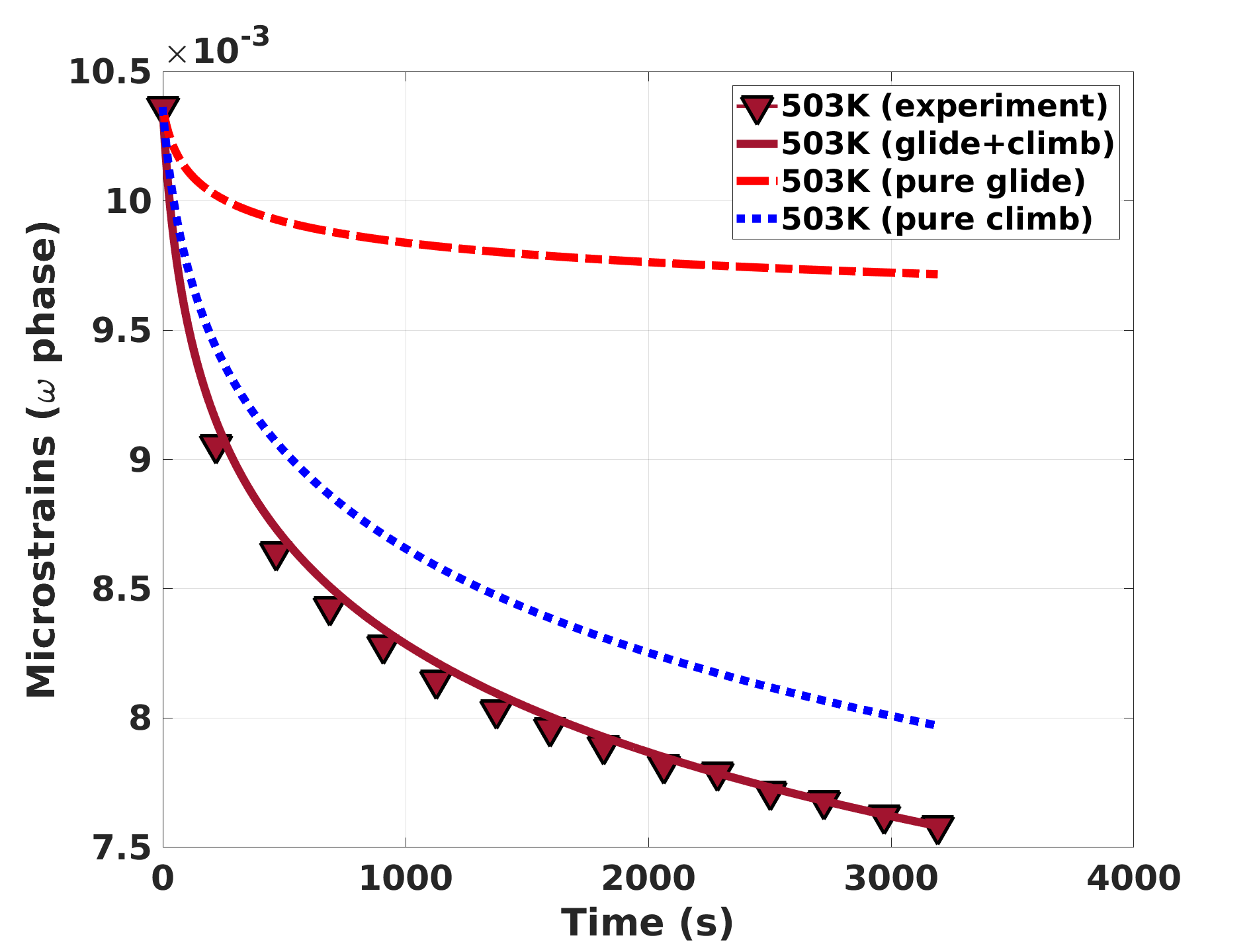}\label{fig:230C_glideclimb}}
	
	\caption{Decomposing the dislocation reduction mechanism into glide and climb contributions for samples heated at 443 K and 503 K. The solid line indicates a combination of both reaction contributions in place whereas the dotted lines indicate the evolution of microstrains under individual contributions.}
	\label{fig:glideclimb}
\end{figure*}

A more detailed picture of the dislocation reduction process is shown in figure \ref{fig:glideclimb}, which illustrates the microstrain evolution in the $\omega$ phase if only one of the two dislocation reduction processes (glide and climb) were included in the model. It can be observed that at lower temperatures (443 K), the glide process dominates strongly in the initial stage but eventually tapers off, in turn letting the climb process dictate dislocation reduction trends at long times. At higher temperatures (503 K), the contribution of climb appears to be more significant and consistently dominates the dislocation removal trends, whereas the red line depicting glide controlled populations sees a rapid exhaustion (figure \ref{fig:230C_glideclimb}). Naturally, the rate of the process rate limited by climb follows an Arrhenius trend with respect to temperature and dominates more at higher temperatures. At sufficient temperatures, if the process is completely climb controlled, standard Avrami kinetics for the transformation should be recovered. Nisoli was able to demonstrate that this indeed occurs for dislocation evolution processes with forms given by equation \ref{eq:nisoli}. Our proposed model and the model of Nisoli et al. both give the same predictions for long times and high temperatures. The key difference is that our model ascribes the behavior to a specific dislocation mechanisms, climb controlled recovery, as opposed to generic thermally activated dislocation motion. 

Since the dislocation reduction trends are coupled to the reverse transformation, the transient behavior of the reverse transformation is associated to the rapid exhaustion of dislocations that eliminate by glide controlled processes. This effect is more obvious at lower temperatures, when the climb controlled processes are more passive and insufficient to drive a rapid elimination of dislocations in the initial stages of annealing. This is also shown in the work of Nisoli et al., in which fits to the $\omega$ decay at low temperatures were unable to capture the initial rapid descent behavior.

It is also important to consider the range of applicability of the current model given our choice to only consider glide and climb controlled recovery mechanisms. Of particular interest is determining if the current framework can be used to assess the $\omega \rightarrow \alpha$ reversion that occurs during the immediate unloading following the shock event. Both dislocation removal mechanisms in the model are essentially controlled by climb of edge segments. Here the assumption is that the time-scale on which individual climb events occurs is short compared to the annealing time and we can consider the rate of occurrence in a statistical sense relative to the total dislocation density. In contrast the shock process provides high strain rates over a period on the order of a  $\mu s$. It is unlikely that dislocation removal via climb leads to the reversion during the shock loading due to the longer time scale under which climb occurs. The current framework works under a longer time length scale \textit{following the shock}, in which the pinning effect of the $\omega$ phase stems from the interactions of dislocation populations \textit{during} the shock. In contrast, the type of plasticity events activated prior to, during, and after phase transformation may vary according to the applied peak shock pressure and duration of shock, implying multiple dislocation interaction possibilities which result in energetically varying post-shock dislocation populations. Unfortunately we do not have a understanding of the material state and defect state during the shock event to speculate on the nature of the immediate $\omega \rightarrow \alpha$ reversion. Modeling the simultaneous reversion would require further knowledge on possible dislocation interactions at high strain rates and how their interaction products eventually pin the $\omega$ phase.  


\subsection{Distinguishing Peak Shock Effects ($\kappa$ parameter)} \label{subsection:kappa2}
Referencing figure \ref{fig:10GPa_simulation}, it was noted that the model was able to generally capture the $\omega$ phase evolution in 10.5 GPa samples despite overpredicting the reduction in microstrain. The simulation results were obtained by directly using parameters calibrated to the 8 GPa samples. We also conjectured that this observational mismatch could be attributed to the difference in defect state in the 10.5 GPa samples. In order to justify this claim, we revisit the VISAR profiles of the shock deformation in figure \ref{fig:visar}, and note that samples shocked to 10.5 GPa complete the $\alpha\to\omega$ transformation in a shorter period of time compared to the samples shocked to 8 GPa. This transformation period is then followed by a period of constant velocity,  indicating continued deformation of the sample at peak pressure during this period. Due to the earlier completion of the transformation, the $\omega$ phase in the 10.5 GPa samples undergoes a longer period under deformation which consequently affords more time for dislocation interactions. Additionally, the exaggerated pressure loading could also activate different slip modes and act to further diversify the interaction possibilities. Ultimately, this could imply that the $\omega$ phase dislocation populations may be distinct, based on the duration of deformation following the $\alpha\to\omega$ transformation. With the exception of the 443K sample, all of the 10.6 GPa samples exhibit the same reduction of microstructure during annealing. This leads us to speculate that there are two distinct populations of defects present: i) the arresting dislocations which are recoverable through the climb and glide processes in our model and ii) dislocations that are not readily removed at these low temperatures but do not exert a significant resisting force on the transformation. 

During the early stages of shock the $\alpha$ phase material undergoes plastic deformation by slip and twinning \cite{cerreta2005influence}, generating a high dislocation density. Following the shear transformation from $\alpha \rightarrow \omega$ a portion of these formerly mobile $\alpha$ dislocations may survive in the $\omega$ in sessile high-energy configurations. We hypothesis that these defects strongly contribute to the arresting force. Given the high energy configuration and high density of these defects, they will rapidly annihilated when they are mobilized. In contrast, we speculate that the second group of dislocations are native to the $\omega$ slip systems and were generated during the longer deformation period during the 10.5 GPa shock.  Kumar and Beyerlein recently computed the generalized stacking fault surfaces for $\omega$-Zr at room temperatures and reported that the favorable slip modes for the $\omega$ phase consist of prismatic $<c>$, prismatic-II $<10\bar{1}0>$ and pyramidal-II $<c+a>$, which are different from the conventional slip modes in the $\alpha$ phase \cite{kumar2017}. As more experimental and computational studies aimed at exploring the fundamental deformation mechanisms of the $\omega$ are performed, we will hopefully be able to verify this hypothesis either by reanalysis of the Low et al. data or by additional in-situ annealing studies. 


In section \ref{sec:jog} we also illustrated the possibility of varying the $\kappa$ parameter to achieve better fits in the 10.5 GPa $\omega$ phase fractions, as shown in figure \ref{fig:kappa}. From a physical understanding, we equated the act of decreasing the $\kappa$ parameter to decreasing the jog spacing in screw dislocations that are removed in a glide controlled manner. Decreasing $\kappa$ effectively describes a more deformed system, as one would expect smaller jog separations in a highly deformed state. From equations \ref{eq:glidegrowth} and \ref{eq:kappa2}, we can see that decreasing the value of $\kappa$ for the 10.5 GPa sample would also decrease the dislocation removal rate, given a constant temperature. One would then logically expect a consequent decrease in the reverse transformation. However, this is not the case as we see in figure \ref{fig:kappa}, in which a lower $\kappa$ value results in a higher initial transformation rate. In order to explain this, a more detailed look at the effect of $\kappa$ on the dislocation evolution and its subsequent effect on the reverse transformation is required. This requires that, we first examine (i) the barrier energy coupling both processes and (ii) the energetic implications of varying $\kappa$. 

\subsubsection{The Barrier Energy}
In previous work, it was observed that the dislocation density in the $\omega$ phase decreased considerably before the $\omega\to\alpha$ transformation occurred in earnest \cite{Brown2014383, low2015}. It was thought that reduction to a certain critical value of defect level in the $\omega$ phase had to be achieved prior to initiation of significant $\omega\to\alpha$ transformation. While sufficient in explaining the rapid initial transformation, this hypothesis was inadequate in explaining the progression of the reverse transformation to new metastable levels depending on the temperature applied. Instead of a critical barrier, this model presents a continuous barrier energy that evolves in tandem with the dislocation density. This barrier energy acts to resist further reverse transformation of the $\omega$ back into $\alpha$. If we associate the immobile dislocations with the arresting of the $\omega$ phase, and that the mobilization and subsequent removal of these dislocations allow for the reverse transformation, we can then alternatively describe the resistive barrier energy as being related to the energy required to mobilize these dislocations. 

\begin{figure}
	\centering
	\includegraphics[scale=0.25]{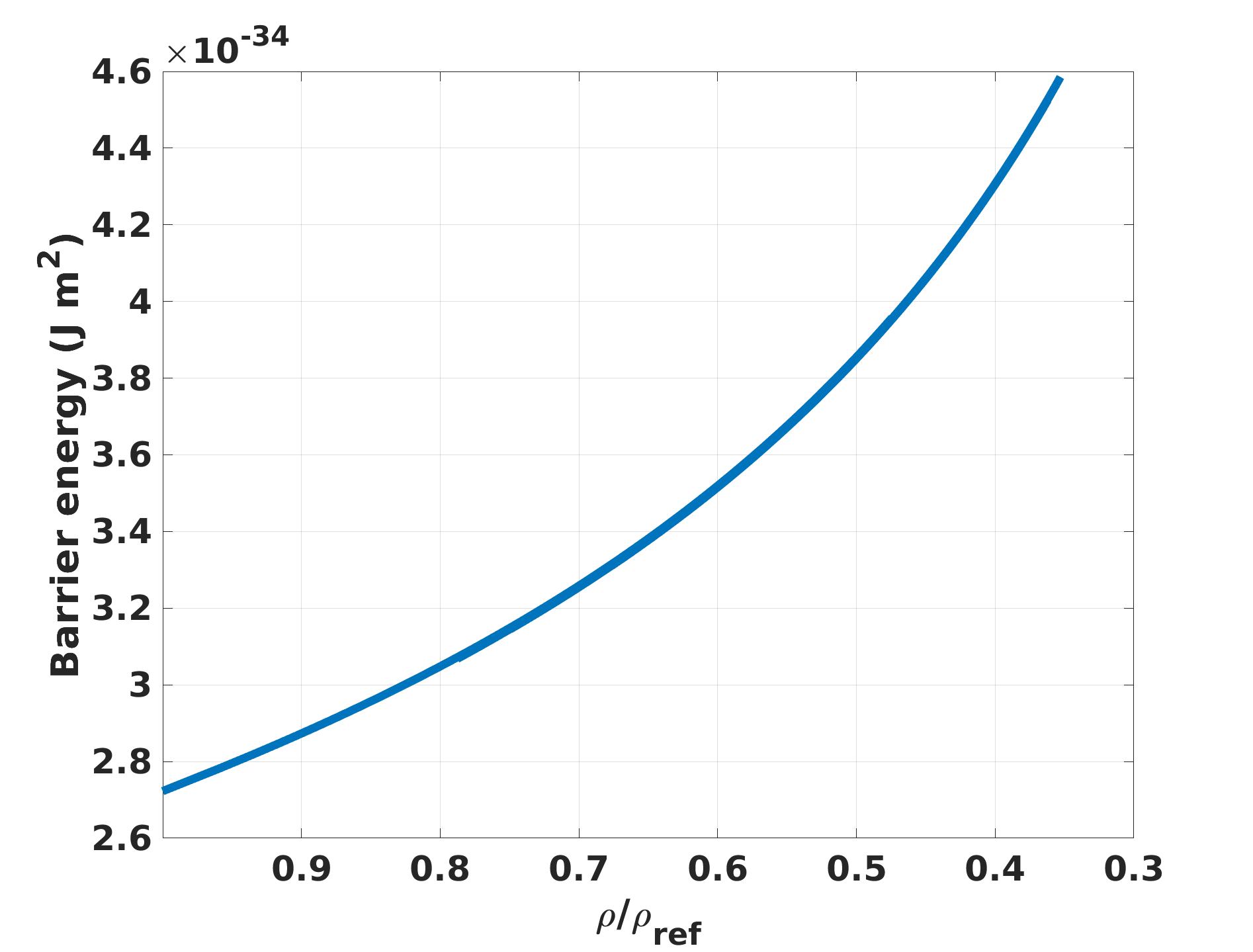}
	\caption{Average barrier energy of individual dislocations in the remaining dislocation population with progression of dislocation removal. Increasing dislocation energies offer increasing resistance to the reverse transformation. The plot was obtained by overlaying multiple plots from the fitting of the isothermal experiments. Here, the dislocation content is normalized by the reference dislocation density $\rho_{ref} = 1.0 \times 10^{15}$ m$^{-2}$.}
	\label{fig:energyprofile}
\end{figure}

This mobilizing energy can be thought to follow a distribution, such that dislocations with lower mobilizing energy requirements are favorably set in motion for elimination reactions. Subsequent removal of dislocations with lower mobilizing energy will mean that the remaining dislocation populations are those with higher mobilizing energy and consequently offer a stronger resistance to the $\omega\rightarrow\alpha$ transformation. In the current model, this mobilizing energy for \textit{individual} dislocations is related to the dislocation energy in a straightforward manner given in equation \ref{eq:barrierenergy}. The increase of the average mobilizing energy with respect to the amount of dislocations remaining is given in figure \ref{fig:energyprofile}. The observation of such an increasing trend falls in line with observed reduction trends in the microstrains (figures \ref{fig:8GPa_omegaStrain} and \ref{fig:10GPa_omegaStrain}), in that a steep reduction occurs in the beginning of the annealing process followed by a much slower reduction in the asymptotic region. The steep reduction in the initial stage comes from the rapid removal of large dislocation populations that are unstable due to their low mobilizing energies (glide controlled), following which they are exhausted and the dislocation removal is proceeded by the removal of more stable dislocation populations (climb controlled). At high homologous temperatures, there is sufficient energy to overcome all barriers and the transformation goes to completion. However, at lower temperatures, a balance is produced between the available thermal energy and the energy necessary to remove dislocations arresting the transformation. In the application to the transformation rate equation (eq. \ref{eq:growthrate}), we note that the barrier energy is normalized by the current dislocation density and represents an averaged resistance of a \textit{single} dislocation at a point in time. Hence, the $\beta_1$ parameter is applied to scale this barrier energy to represent a collective contribution from an effective dislocation population. In the current implementation, we have chosen $\beta_1 = 1.9\times 10^{14}$ which is $\sim$ 20\% of the reference dislocation density $\rho_{ref}=1 \times 10^{15}$. 

\subsubsection{Varying the $\kappa$ Parameter}
From figure \ref{fig:kappa}, decreasing the value of $\kappa$ was shown to increase the initial transformation rate, although inspection of equations \ref{eq:glidegrowth} and \ref{eq:kappa2} indicate that it should lower the removal rate of glide dislocation populations and subsequently reduce the initial transformation rate. The explanation for this observation is that the behavior of the reverse transformation is not purely reliant on dislocation removal, but rather the energetics of the removal process. In the current model, the barrier energy resisting the $\omega\to\alpha$ transformation is described as being related to the mobilizing energy requirement of the dislocation populations. By decreasing the $\kappa$ parameter, we are also effectively increasing the mobilizing energy requirement for the glide dislocation populations, and consequently slowing down the dislocation recovery process. Equivalently, we are widening the energy distribution \textit{solely} for the glide populations while keeping the same dislocation content so that the glide population now includes dislocations with higher energy requirements for mobilization. However, this energy increase does not shift the glide populations to be on par with the climb populations, they are still exhausted before climb controlled recovery. Since the mobilization energy requirement of a dislocation is directly related to its resistive strength against the transformation, in the case of the inevitable exhaustion of glide populations, increasing their energetic requirements increases their resistive strengths against the transformation, and that removal of a population of `stronger' resistive dislocations results in a faster initial reverse transformation. To simplify this argument further, varying $\kappa$ effectively determines how strongly the glide population arrests the reverse transformation. We see this reflected in figure \ref{fig:kappa}, in which decreasing the jog spacing $\kappa$ results in a higher initial rate of transformation, with minor effects on the asymptotic trend.

\begin{figure}
	\centering
	\includegraphics[scale=0.25]{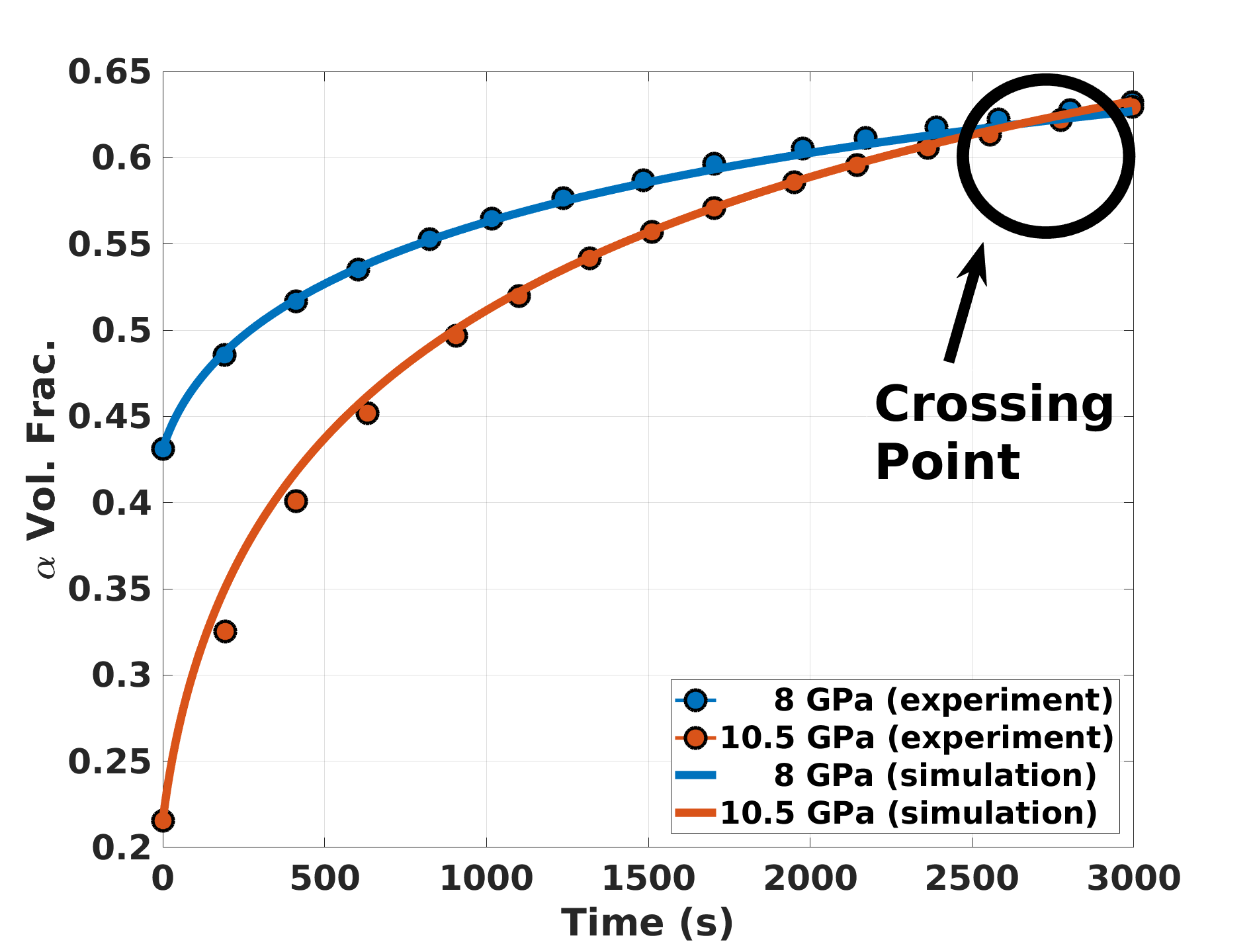}
	\caption{Experimental data and simulation fits to compare the effects of peak pressure on the reverse transformation at 483 K.}
	\label{fig:pressurecompare}
\end{figure}

\begin{figure}
	\centering
	\includegraphics[scale=0.25]{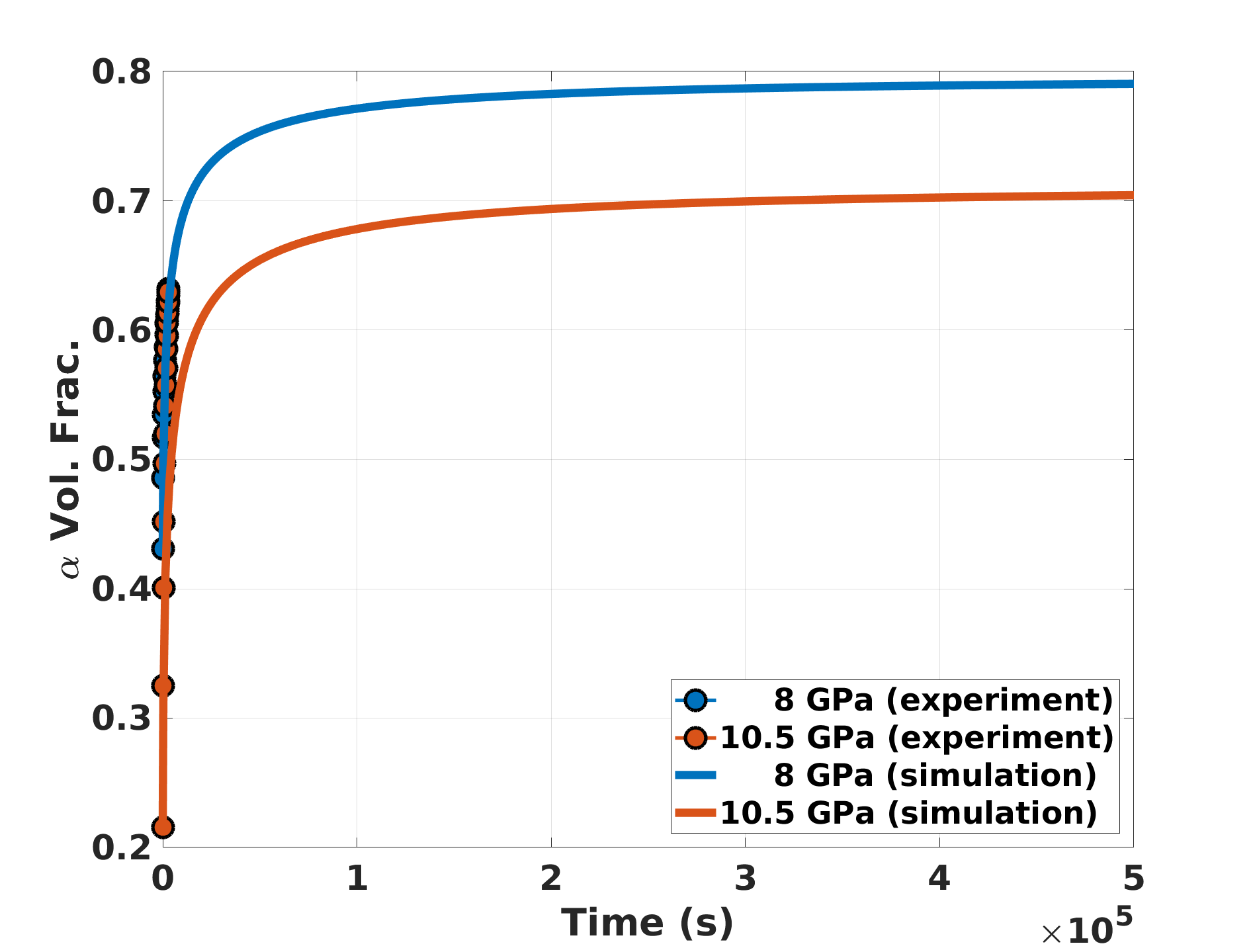}
	\caption{Hypothetical simulation of samples annealed at 483 K to longer times. The $\kappa$ parameter is set equal ($\kappa$=0.023) for both cases to indicate similar post shock defect states. An absence of cross-over behavior is observed.  }
	\label{fig:pressurecompare_asymptotic}
\end{figure}

In the isothermal annealing experiments, given sufficiently high temperatures ($>483$ K), a cross-over trend was observed when comparing the $\alpha$ phase fractions of both the 8 and 10.5 GPa samples. More specifically, even though the 10.5 GPa samples start off with a lower amount of $\alpha$ phase than the 8 GPa samples, the growth of $\alpha$ in the 10.5 GPa is rapid enough that it catches up and surpasses the $\alpha$ content in the 8 GPa samples. Figure \ref{fig:pressurecompare} depicts the cross-over trend observed at 483 K when comparing samples shocked different pressures. Note that the fit applied to the 10.5 GPa curve here is the same one that was used in figure \ref{fig:kappa} with $\kappa=0.0215$. Low et al. speculated that the transformation rate was directly related to the current $\omega$ fraction so that similar cross-over behavior would be observed anytime two samples with different initial $\omega$ content were simultaneously annealed given enough time. In the model analysis, we presented an alternate view that the transformation rate was more related to the types of defects present and not the current $\omega$ fraction.  In order to highlight the difference in these views we ran an additional simulation where both the 8 and 10.5 GPa samples were assigned equal values of $\kappa=0.023$ and annealed at 483 K to longer times of $t=100,000$ s and found no cross-over points. Figure \ref{fig:pressurecompare_asymptotic} illustrates the results of this hypothetical simulation. Notice that at asymptotic times, both 8 and 10.5 GPa samples have achieved a plateau in the reverse transformation, and are unlikely to cross-over at further times. Again, this result points to the difference in glide controlled dislocation population as the microstructural feature which directly controls the initial transient transformation rate. This argument is further solidified by the observation that at asymptotic times, which we previously established to be dictated by climb controlled removal of dislocations, both samples behave identically and are metastable to long times with no observable trends pointing to a cross-over. 

\subsection{Nucleation vs. Interfacial Growth of $\alpha$} \label{sec:discuss_alpha1}
As discussed earlier, the model is able to be extrapolated across a range of temperatures to reproduce Brown et al.'s annealing experiments as shown in figure \ref{fig:constantRamp}. Upon close observation of the simulation prediction in figure \ref{fig:constantRamp}, the volume fraction curves have abrupt changes in the slope (which are especially noticeable in the beginning of the transformation). This is related to the linear piecewise fitting that has been applied to the $\alpha_1$ parameter as depicted in figure \ref{fig:alpha1}. Here, we attempt to ascribe a meaning to the $\alpha_1$ parameter. Referring to figure \ref{fig:alpha1}, if we consider only the temperature range from the isothermal experiments, and proceed along the x-axis in increasing temperature, we would observe a rapid initial rise of the $\alpha_1$ parameter (stage 1) followed by a very noticeable decrease in the slope at $ 463 K<\theta<483 K$ (stage 2). Further increasing the temperature sees another rapid increase in the $\alpha_1$ parameter (stage 3), although slightly less rapid as the initial increase. 

In Low et al. the authors compared the EBSD scans of samples in the as shocked state, post annealing at 443 K, and post annealing at 463 K. In the sample annealed at 443 K, the growth of the $\alpha$ phase seemed to proceed by interfacial growth of existing $\alpha$ lath as the number of laths in the annealed condition did not increase but the laths were visibly thicker \cite{low2015}. In the sample that was annealed at 463 K, the volume fraction of the $\alpha$ phase was not only higher but the number of $\alpha$ laths increased and the average lath was significantly thinner suggesting nucleation and growth of new laths. Here, we suggest that the temperature trends of the $\alpha_1$ parameter reflect the transition of the reverse transformation from being interfacial growth dominant to being nucleation dominant, and that this transition occurs in stage 2 of the $\alpha_1$ vs. temperature fitting. Although the current model supports the hypothesis of the reverse transformation being closely related to the removal of defects, much work is left to be done in terms of understanding the mechanisms by which the reverse transformation occurs, eg. the type of defects at the interface and their removal mechanisms which lead to growth of current interfaces or nucleation of a new $\alpha$ lath. For the purposes of this model the $\alpha_1$ parameter then acts as a modifier to the bulk Gibbs free energy difference to account for the variation in specific transformation mechanisms active at a given temperature.  

\section{Conclusion}
In prior work, in-situ XRD experiments were undertaken to investigate the thermal stability of the $\alpha/\omega$ microstructure generated from shock loading pure Zr. Provided sufficient thermal energy, the material exhibited a rapid initial transformation rate back to $\alpha$ followed by a plateau to a metastable state with a lower amount of retained $\omega$. Similar trends were observed in the dislocation densities (associated with the measured microstrains) in the $\omega$ phase. However, comparing both reductions simultaneously revealed that a significant reduction of dislocation densities in the $\omega$ phase preceded the reverse transformation, leading to postulation that the dislocation populations within the $\omega$ phase were responsible for the metastability of the $\omega$ phase at ambient conditions. Building atop this hypothesis, a model that couples the dislocation removal to the reverse transformation has been developed. Both phenomena are coupled together via a barrier energy that is a function of the defect state of the microstructure i.e. the remaining dislocation population that effectively resists the reverse transformation. The removal of dislocations favors of defects that have a lower energetic requirement for mobilization. As more dislocations with low mobilizing energies are removed, the remaining dislocation population are much more stable and harder to mobilize, effectively providing a stronger resistance to the reverse transformation, thus the occurrence of a plateau to metastable states with retained $\alpha$ under heating.

The dislocation removal in the $\omega$ phase has been attributed to two distinct modes of occurrence, namely the glide and climb controlled process. Glide controlled removal of dislocations has been demonstrated to dictate or aid significantly in the initial, rapid reduction of dislocations. Upon exhaustion of dislocations population associated with glide elimination, the asymptotic rate removal then becomes climb controlled. Due to the coupling between dislocation reduction and the reverse transformation, the association of the dislocation reduction with two simultaneously occurring modes (which differ in contribution strength depending on the annealing stage) allows for a better and sensible description of the transient and asymptotic plateauing of the reverse transformation.

The model also adequately describes the effects of peak shock pressures (difference in extent of deformation) on the annealing responses. This is done via the $\kappa$ parameter from equation \ref{eq:glidegrowth} which relates the jog separation on screw dislocations to the average dislocation separation distance. A lower $\kappa$ value reflects a more deformed state of dislocations. Doing so allowed for verification of cross-over trends observed in annealing the experiments, in which the 10.5 GPa samples ended up with lower amounts of retained $\omega$ compared to 8 GPa samples under similar isothermal environments, despite the 10.5 GPa samples starting off with higher amounts of $\omega$.

\section*{References}

\bibliography{biblio_list}

\end{document}